\pdfoutput=1
\documentclass[12pt,a4paper]{article}

\usepackage{ifthen} \usepackage{comment}
\usepackage{float}
\newboolean{pdflatex}
\setboolean{pdflatex}{true} 

\newboolean{articletitles}
\setboolean{articletitles}{true} \usepackage{siunitx}
\newboolean{uprightparticles}
\setboolean{uprightparticles}{false} 

\newcolumntype{R}{>{}r<{}r<{}} \newcolumntype{L}{>{}l<{}l<{}}

\def\paperauthors{LHCb collaboration} \def\paperasciititle{Search for CP violation using T-odd correlations in B0 -> p pbar K+ pi- decays} \def\papertitle{Search for $C\!P$ violation using $\hat{T}$-odd correlations in \\  $\B^{0} \to p
\bar p K^{+} \pi^{-}$ decays} \def\paperkeywords{{High Energy Physics}, {LHCb}} \def\papercopyright{\the\year\ CERN for the benefit of the LHCb collaboration} \def\paperlicence{CC BY 4.0 licence}
\def\paperlicenceurl{https://creativecommons.org/licenses/by/4.0/}

\usepackage[top=1in, bottom=1.25in, left=1in, right=1in]{geometry}

\usepackage{microtype}
\usepackage{lineno}  \usepackage{xspace} \usepackage{caption}

\usepackage{graphicx}  \usepackage{color}
\usepackage{colortbl}
\graphicspath{{./figs/}} 

\usepackage{amsmath} \usepackage{amssymb}
\usepackage{amsfonts}
\usepackage{upgreek} 

\newcommand*\patchAmsMathEnvironmentForLineno[1]{\expandafter\let\csname old#1\expandafter\endcsname\csname #1\endcsname
\expandafter\let\csname oldend#1\expandafter\endcsname\csname
end#1\endcsname
 \renewenvironment{#1}{\linenomath\csname old#1\endcsname}{\csname oldend#1\endcsname\endlinenomath}}
\newcommand*\patchBothAmsMathEnvironmentsForLineno[1]{\patchAmsMathEnvironmentForLineno{#1}\patchAmsMathEnvironmentForLineno{#1*}}
\AtBeginDocument{\patchBothAmsMathEnvironmentsForLineno{equation}\patchBothAmsMathEnvironmentsForLineno{align}\patchBothAmsMathEnvironmentsForLineno{flalign}\patchBothAmsMathEnvironmentsForLineno{alignat}\patchBothAmsMathEnvironmentsForLineno{gather}\patchBothAmsMathEnvironmentsForLineno{multline}\patchBothAmsMathEnvironmentsForLineno{eqnarray}}

\usepackage{hyperxmp}

\usepackage[pdftex,
            pdfauthor={\paperauthors},
            pdftitle={\paperasciititle},
            pdfkeywords={\paperkeywords},
            pdfcopyright={Copyright (C) \papercopyright},
            pdflicenseurl={\paperlicenceurl}]{hyperref}

\usepackage[colorinlistoftodos,textsize=scriptsize]{todonotes}

\usepackage[bottom,flushmargin,hang,multiple]{footmisc}

\usepackage[all]{hypcap} 

\usepackage{xspace} 
\usepackage{upgreek}

\def\lhcb   {\mbox{LHCb}\xspace}

\def\MagUp {\mbox{\em Mag\kern -0.05em Up}\xspace}

\ifthenelse{\boolean{uprightparticles}}{

 \def\PDelta      {\ensuremath{\Delta}\xspace}                 
 \def\PXi         {\ensuremath{\Xi}\xspace}                 
 \def\PLambda     {\ensuremath{\Lambda}\xspace}                 
 \def\PSigma      {\ensuremath{\Sigma}\xspace}                 
 \def\POmega      {\ensuremath{\Omega}\xspace}                 
 \def\PUpsilon    {\ensuremath{\Upsilon}\xspace}
 \let\oldPi\Pi
 \def\PPi         {\ensuremath{\oldPi}\xspace}

 \def\PB      {\ensuremath{\mathrm{B}}\xspace}                 
                  
 \def\PD      {\ensuremath{\mathrm{D}}\xspace}

 \def\PK      {\ensuremath{\mathrm{K}}\xspace}

 \def\Pb      {\ensuremath{\mathrm{b}}\xspace}                 
 \def\Pc      {\ensuremath{\mathrm{c}}\xspace}

 \def\Pi      {\ensuremath{\mathrm{i}}\xspace}

 \def\Ps      {\ensuremath{\mathrm{s}}\xspace}

 \def\thebaroffset{0.0em}
}
{

 \mathchardef\PDelta="7101
 \mathchardef\PXi="7104
 \mathchardef\PLambda="7103
 \mathchardef\PSigma="7106
 \mathchardef\POmega="710A
 \mathchardef\PUpsilon="7107
 \mathchardef\PPi="7105
                  
 \def\PB      {\ensuremath{B}\xspace}                 
                  
 \def\PD      {\ensuremath{D}\xspace}

 \def\PK      {\ensuremath{K}\xspace}

 \def\Pb      {\ensuremath{b}\xspace}                 
 \def\Pc      {\ensuremath{c}\xspace}

 \def\Pi      {\ensuremath{i}\xspace}

 \def\Ps      {\ensuremath{s}\xspace}

 \def\thebaroffset{0.18em}
}
\newcommand{\offsetoverline}[2][\thebaroffset]{\kern #1\overline{\kern -#1 #2}}

\makeatletter
\ifcase \@ptsize \relax \newcommand{\miniscule}{\@setfontsize\miniscule{4}{5}}\or \newcommand{\miniscule}{\@setfontsize\miniscule{5}{6}}\or \newcommand{\miniscule}{\@setfontsize\miniscule{5}{6}}\fi
\makeatother

\DeclareRobustCommand{\optbar}[1]{\shortstack{{\miniscule (\rule[.5ex]{1.25em}{.18mm})}
  \\ [-.7ex] $#1$}}

\def\squark    {{\ensuremath{\Ps}}\xspace}

\def\cquark    {{\ensuremath{\Pc}}\xspace}

\def\bquark    {{\ensuremath{\Pb}}\xspace}

\def\KorKbar {\kern \thebaroffset\optbar{\kern -\thebaroffset \PK}{}\xspace}

\def\Dbar    {{\ensuremath{\offsetoverline{\PD}}}\xspace}
\def\D       {{\ensuremath{\PD}}\xspace}

\def\DorDbar {\kern \thebaroffset\optbar{\kern -\thebaroffset \PD}\xspace}

\def\Dzb     {{\ensuremath{\Dbar{}^0}}\xspace}
\def\Dp      {{\ensuremath{\D^+}}\xspace}
\def\Dm      {{\ensuremath{\D^-}}\xspace}

\def\DpDm    {\ensuremath{\Dp {\kern -0.16em \Dm}}\xspace}

\def\B       {{\ensuremath{\PB}}\xspace}

\def\BorBbar {\kern \thebaroffset\optbar{\kern -\thebaroffset \PB}\xspace}

\def\Bd      {{\ensuremath{\B^0}}\xspace}

\def\BdorBdbar {\kern \thebaroffset\optbar{\kern -\thebaroffset \Bd}\xspace}

\def\Bs      {{\ensuremath{\B^0_\squark}}\xspace}

\def\BsorBsbar {\kern \thebaroffset\optbar{\kern -\thebaroffset \Bs}\xspace}

\def\Y#1S{\ensuremath{\PUpsilon{(#1S)}}\xspace}

\def\Lbar        {{\ensuremath{\offsetoverline{\PLambda}}}\xspace}
\def\LorLbar     {\kern \thebaroffset\optbar{\kern -\thebaroffset \PLambda}\xspace}

\def\Lcbar       {{\ensuremath{\Lbar{}^-_\cquark}}\xspace}

\def\to                 {\ensuremath{\rightarrow}\xspace}

\def\CP                {{\ensuremath{C\!P}}\xspace}

\def\AT#1     {\ensuremath{A_{\mathrm{T}}^{#1}}\xspace}

\def\C#1      {\ensuremath{\mathcal{C}_{#1}}\xspace}                       \def\Cp#1     {\ensuremath{\mathcal{C}_{#1}^{'}}\xspace}                    \def\Ceff#1   {\ensuremath{\mathcal{C}_{#1}^{\mathrm{(eff)}}}\xspace}        \def\Cpeff#1  {\ensuremath{\mathcal{C}_{#1}^{'\mathrm{(eff)}}}\xspace}       \def\Ope#1    {\ensuremath{\mathcal{O}_{#1}}\xspace}                       \def\Opep#1   {\ensuremath{\mathcal{O}_{#1}^{'}}\xspace}

\newcommand{\nospaceunit}[1]{\ensuremath{\text{#1}}}       
\newcommand{\aunit}[1]{\ensuremath{\text{\,#1}}}

\newcommand{\tev}{\aunit{Te\kern -0.1em V}\xspace}
\newcommand{\gev}{\aunit{Ge\kern -0.1em V}\xspace}
\newcommand{\mev}{\aunit{Me\kern -0.1em V}\xspace}
\newcommand{\kev}{\aunit{ke\kern -0.1em V}\xspace}
\newcommand{\ev}{\aunit{e\kern -0.1em V}\xspace}
 
\newcommand{\mevc}{\ensuremath{\aunit{Me\kern -0.1em V\!/}c}\xspace}
\newcommand{\gevc}{\ensuremath{\aunit{Ge\kern -0.1em V\!/}c}\xspace}
\newcommand{\mevcc}{\ensuremath{\aunit{Me\kern -0.1em V\!/}c^2}\xspace}
\newcommand{\gevcc}{\ensuremath{\aunit{Ge\kern -0.1em V\!/}c^2}\xspace}

\def\mum  {\ensuremath{\,\upmu\nospaceunit{m}}\xspace}

\def\fb   {\ensuremath{\aunit{fb}}\xspace}
\def\invfb   {\ensuremath{\fb^{-1}}\xspace}

\newcommand{\chisqip}{\ensuremath{\chi^2_{\text{IP}}}\xspace}

\def\gsim{{~\raise.15em\hbox{$>$}\kern-.85em
          \lower.35em\hbox{$\sim$}~}\xspace}
\def\lsim{{~\raise.15em\hbox{$<$}\kern-.85em
          \lower.35em\hbox{$\sim$}~}\xspace}

\def\pt         {\ensuremath{p_{\mathrm{T}}}\xspace}

\def\ptot       {\ensuremath{p}\xspace}

\def\evtgen     {\mbox{\textsc{EvtGen}}\xspace}

\def\geant      {\mbox{\textsc{Geant4}}\xspace}

\def\photos     {\mbox{\textsc{Photos}}\xspace}

\def\pythia     {\mbox{\textsc{Pythia}}\xspace}

\def\tell1  {TELL1\xspace}
\def\ukl1   {UKL1\xspace}

\newcommand{\lhcborcid}[1]{\href{https://orcid.org/#1}{\hspace*{0.1em}\raisebox{-0.45ex}{\includegraphics[width=1em]{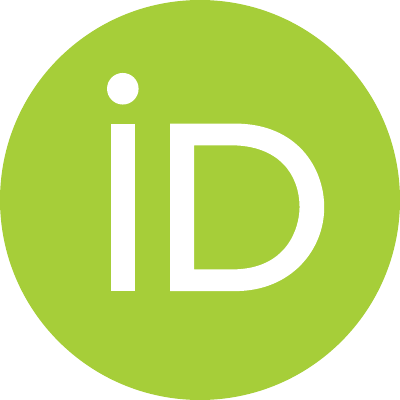}}}}

\usepackage{cite} \usepackage{mciteplus}
 \usepackage{longtable} 

\begin{document}

\renewcommand{\thefootnote}{\fnsymbol{footnote}}
\setcounter{footnote}{1}

\begin{titlepage}
\pagenumbering{roman}

\vspace*{-1.5cm}
\centerline{\large EUROPEAN ORGANIZATION FOR NUCLEAR RESEARCH (CERN)}
\vspace*{1.5cm}
\noindent
\begin{tabular*}{\linewidth}{lc@{\extracolsep{\fill}}r@{\extracolsep{0pt}}}
\ifthenelse{\boolean{pdflatex}}{\vspace*{-1.5cm}\mbox{\!\!\!\includegraphics[width=.14\textwidth]{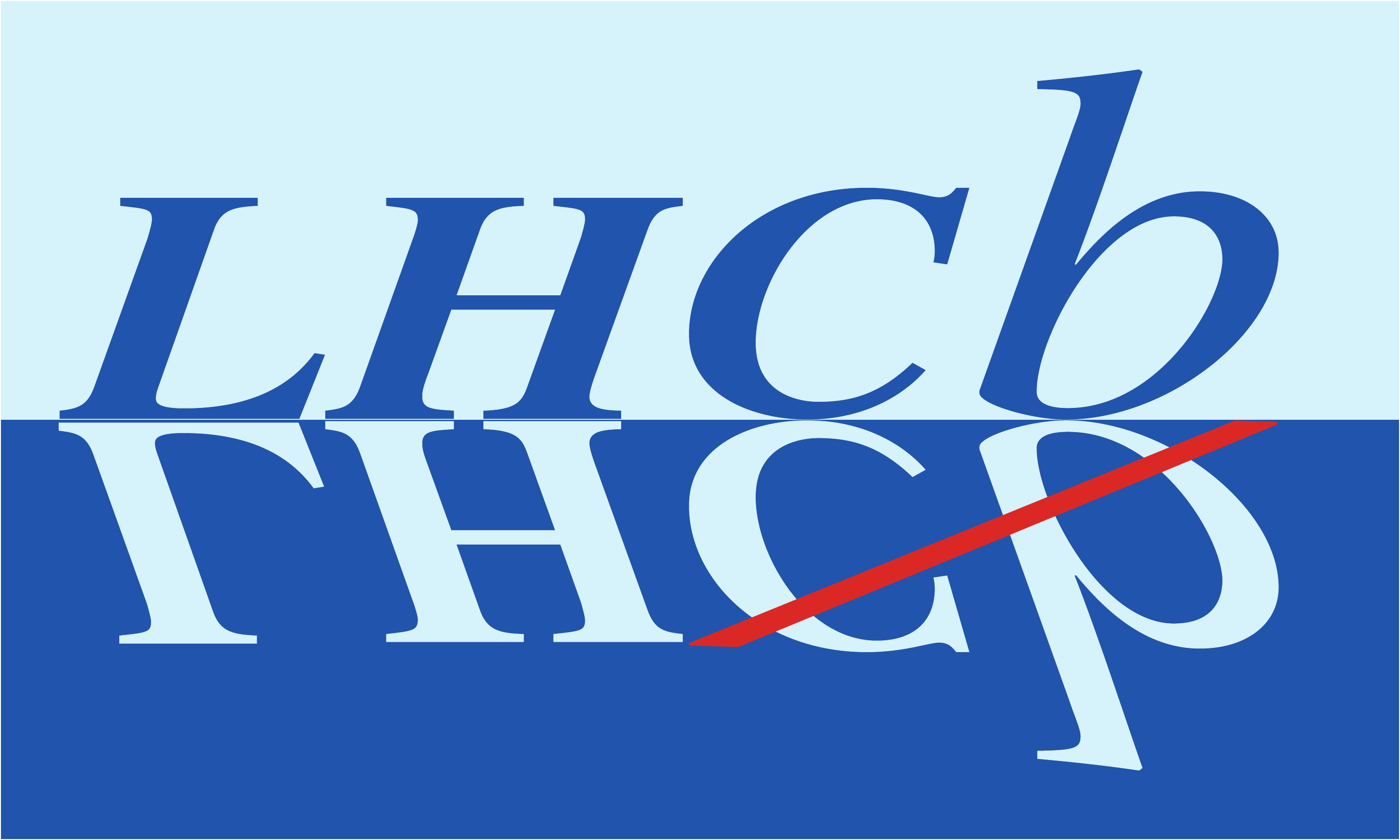}} & &}{\vspace*{-1.2cm}\mbox{\!\!\!\includegraphics[width=.12\textwidth]{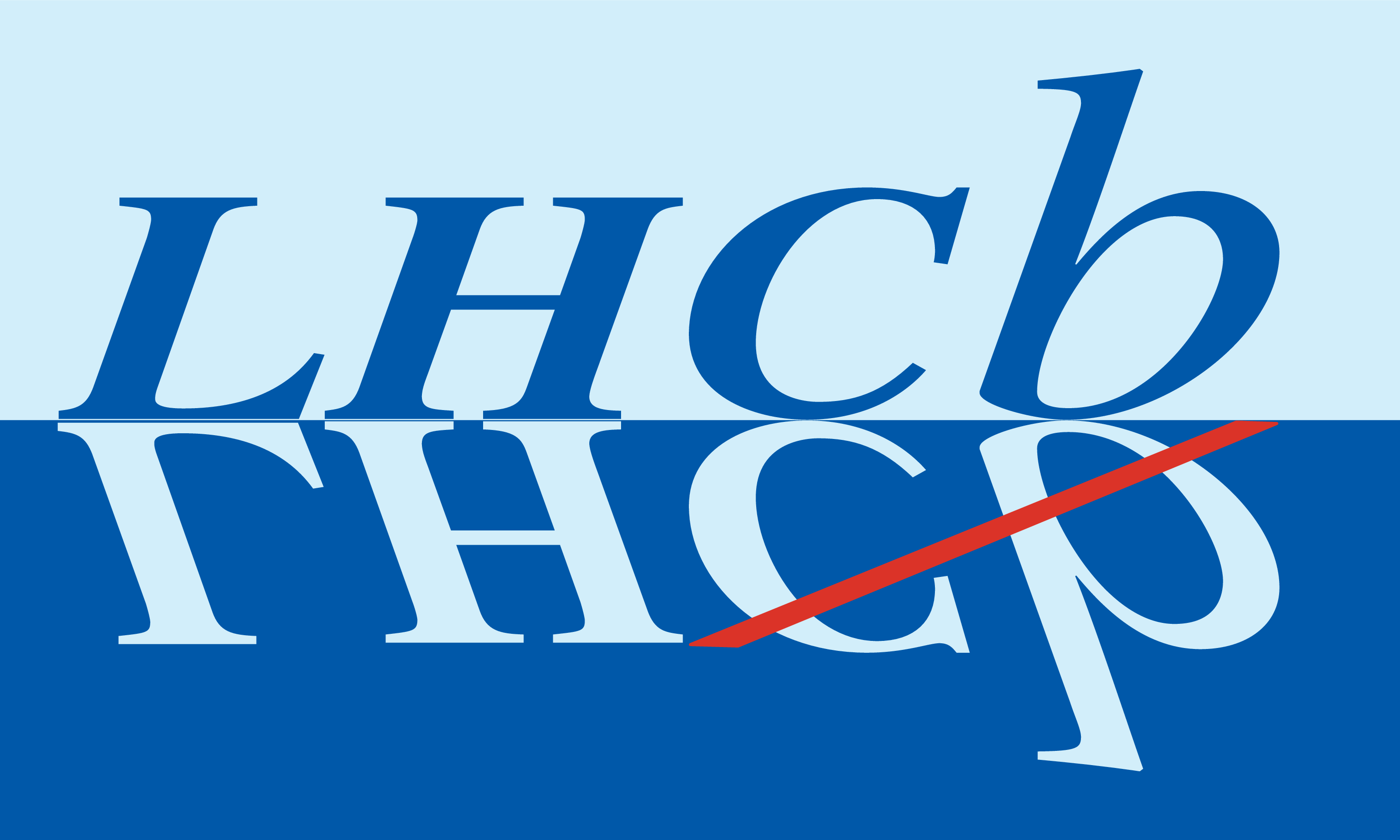}} & &}\\
 & & CERN-EP-2022-083 \\  & & LHCb-PAPER-2022-003 \\  & & \today \\ & & \\
\end{tabular*}

\vspace*{4.0cm}

{\normalfont\bfseries\boldmath\huge
\begin{center}
\papertitle 
\end{center}
}

\vspace*{2.0cm}

\begin{center}
\paperauthors\footnote{Authors are listed at the end of this Letter.}
\end{center}

\vspace{\fill}

\begin{abstract}
  \noindent
  A search for \CP and $P$ violation in charmless four-body
$B^{0} \to p \bar p K^{+} \pi^{-}$ decays is performed using triple-product asymmetry observables.
It is based on 
proton-proton collision data collected by the LHCb experiment at centre-of-mass energies of $7$, $8$ and $13$\tev, corresponding to a total integrated luminosity of $8.4\invfb$. The \CP- and $P$-violating asymmetries are measured both in the integrated phase space and in specific regions. No evidence is seen for \CP violation. $P$-parity violation is observed at a significance of 5.8 standard deviations.

\end{abstract}

\vspace*{2.0cm}

\begin{center}
  Published in
  Phys.~Rev.~D108 (2023) 032007
\end{center}

\vspace{\fill}

{\footnotesize 
\centerline{\copyright~\papercopyright. \href{\paperlicenceurl}{\paperlicence}.}}
\vspace*{2mm}

\end{titlepage}

\newpage
\setcounter{page}{2}
\mbox{~}

\renewcommand{\thefootnote}{\arabic{footnote}}
\setcounter{footnote}{0}

\cleardoublepage

\pagestyle{plain} \setcounter{page}{1}
\pagenumbering{arabic}


\section{Introduction}
\label{sec:IntroductionB2ppbarKpi}

Studying \CP violation in $b$-hadron decays is one of the main purposes of
the \lhcb experiment, aimed at testing the validity of the Cabibbo-Kobayashi-Maskawa (CKM) mechanism in the Standard Model (SM). New sources of \CP violation, beyond the CKM mechanism, can provide insights into the matter-antimatter
asymmetry observed in the universe. Multi-body $B$-meson decays have proven to be an excellent laboratory for studying \CP violation thanks to significant interference between the underlying amplitudes. Indeed, large \CP asymmetries localised in regions of phase space of charmless three-body
$B$-meson decays have been reported by the \lhcb collaboration~\cite{LHCb-PAPER-2019-017, LHCb-PAPER-2018-051, LHCB-PAPER-2013-051, LHCB-PAPER-2013-027}, including the first evidence of \CP violation in the $B^{+}\to p\bar{p} K^{+}$ decay~\cite{LHCb-PAPER-2014-034}. It is therefore of great interest to search for further manifestations of \CP violation in baryonic $B$ decays, where asymmetries of up to 20\% are predicted~\cite{hsiao, PhysRevLett.98.011801, hsiao2}.

In this paper, a search for \CP and $P$ violation based on triple-product asymmetries~\cite{PhysRevD.92.076013} in the charmless region of the $B^{0} \to p\bar{p} K^{+}\pi^{-}$ decay\footnote{Charge-conjugated decays are implicitly considered throughout the text.} is reported using proton-proton ($pp$) collision data collected with the LHCb detector, corresponding to a total integrated luminosity of $8.4\invfb$. A data subsample
of $3\invfb$ was
collected at centre-of-mass energies
of 7 and 8 TeV during 2011 and 2012 (denoted Run 1) while a data subsample of $5.4\invfb$
was collected at 13 TeV
from 2016 to 2018 (denoted Run 2).

\begin{figure}[H]
    \centering
    \includegraphics[scale=0.25]{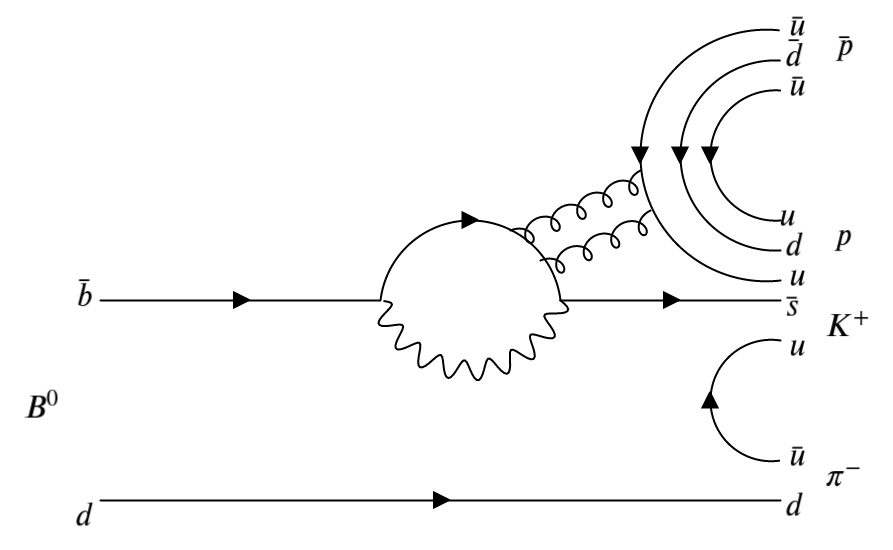}
    \includegraphics[scale=0.25]{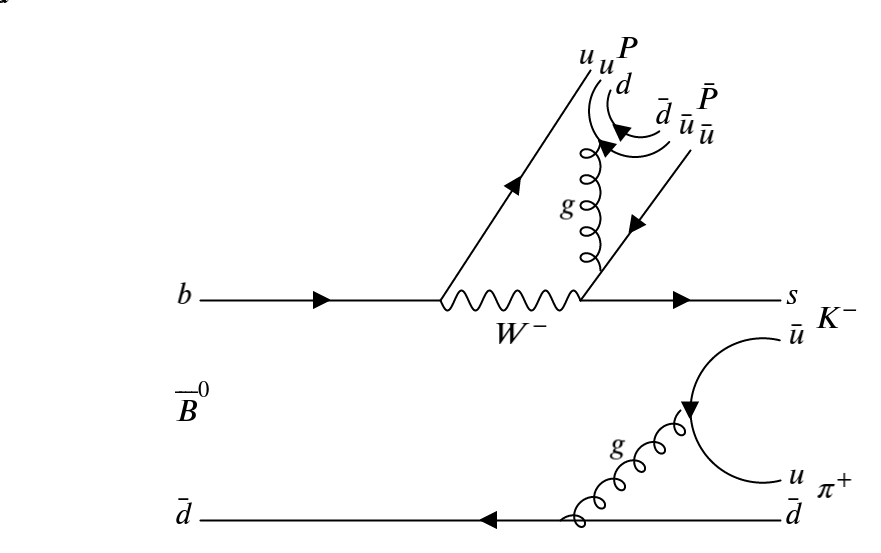}
    \caption{Feynman diagrams for the $B^{0}\rightarrow p\bar{p}K^{+}\pi^{-}$ decay in the charmless region: penguin (left) and tree (right). }
    \label{fig:feynmandiagram}
\end{figure}

The study is performed for proton-antiproton invariant mass $m_{p\bar{p}} < 2.85$\gevcc, corresponding to a region below the charmonium resonances. 
In this region, the decay is governed mainly by tree-level $b\to u \bar{u} s$ and loop-level $b\to s \bar{u} u$ transitions, as shown in Fig.~\ref{fig:feynmandiagram}. 

 Violation of the \CP symmetry can arise from the interference of these two amplitudes, whose weak-phase difference is given by arg$(V_{ub}V^{*}_{us}/V_{tb}V^{*}_{ts})$, and is approximately equal to the CKM angle $\gamma$ in the SM~\cite{TPAinLambdab}. However, it is noted that this work is largely exploratory since no precise SM prediction is available yet for the full phase space of the channel under study. The only prediction available refers to the $B^{0} \to p \bar{p} K^{*0}$ decay channel with an expected 1\% asymmetry~\cite{hsiao}. 
 
The three-momenta of the final-state particles in the $B^{0}$ and $\bar B^{0}$
rest frame are used to build the triple-products $C_{\hat{T}}$ for $B^{0}$ and $\bar C_{\hat{T}}$ for $\bar B^{0}$, which are odd under the operator $\hat{T}$ that reverses the momentum of the particles, and thus acts similarly to the $P$-parity operator. These triple products are
defined as
\begin{equation}
C_{\hat{T}} = \vec{p}_{K^{+}} \cdot (\vec{p}_{\pi^{-}} \times \vec{p}_{p}), \quad 
{\bar C}_{\hat{T}} = \vec{p}_{K^{-}} \cdot (\vec{p}_{\pi^{+}} \times \vec{p}_{\bar p}).
\end{equation}
where $\vec{p}$ denotes vector momentum of the final-state particle indicated in the subscript.
Under the $\CP$ operator the triple product transforms as $\CP(C_{\hat{T}})=-\bar{C}_{\hat{T}}$.

The two $\hat{T}$-odd triple product asymmetries are defined as~\cite{Datta:2003mj}
\begin{equation}
A_{\hat{T}} = \frac{N(C_{\hat{T}}>0)-N(C_{\hat{T}}<0)}{N(C_{\hat{T}}>0)+N(C_{\hat{T}}<0)},
\quad  \bar{A}_{\hat{T}} = \frac{\bar{N}(-\bar{C}_{\hat{T}}>0)-\bar{N}(-\bar{C}_{\hat{T}}<0)}{\bar{N}(-\bar{C}_{\hat{T}}>0)+\bar{N}(-\bar{C}_{\hat{T}}<0)},
\end{equation}
where $N$ and $\bar{N}$ are the numbers of $B^{0}$ and $\bar B^{0}$
decays satisfying the requirement expressed in the corresponding parenthesis.
The \CP- and $P$-violating observables are then constructed as~\cite{Datta:2003mj}
\begin{equation}
a^{\hat{T}\text{-odd}}_{\CP} = \frac{1}{2}(A_{\hat{T}}-\bar A_{\hat{T}}), \quad a^{\hat{T}\text{-odd}}_{P} = \frac{1}{2}(A_{\hat{T}}+\bar A_{\hat{T}}).
\label{eq:atodd}
\end{equation}
A significant deviation from zero in these two observables would indicate \CP violation and $P$ violation, respectively. In contrast to the asymmetry between the phase-space
integrated rates, triple-product asymmetries are sensitive to the interference of $\hat{P}$-even and \mbox{$\hat{P}$-odd} amplitudes
and thus have a different sensitivity to strong phases \cite{PhysRevD.92.076013,PhysRevD.39.3339}.
Triple-product asymmetries have been used to search for \CP violation in \bquark-baryon decays~\cite{LHCb-PAPER-2018-001, LHCb-PAPER-2019-028} and in $D$-meson decays~\cite{LHCb-PAPER-2014-046,LHCb-PAPER-2016-044}. 
By construction, such asymmetries are largely insensitive to particle-antiparticle production and detector-induced asymmetries~\cite{LHCb-PAPER-2014-046}.  

\section{Detector and simulation}
\label{sec:Detector}

The \lhcb detector~\cite{LHCb-DP-2008-001,LHCb-DP-2014-002} is a single-arm forward
spectrometer covering the \mbox{pseudorapidity} range $2<\eta <5$,
designed for the study of particles containing \bquark or \cquark
quarks. The detector includes a high-precision tracking system
consisting of a silicon-strip vertex detector surrounding the $pp$
interaction region~\cite{LHCb-DP-2014-001}, a large-area silicon-strip detector located
upstream of a dipole magnet with a bending power of about
$4{\mathrm{\,Tm}}$, and three stations of silicon-strip detectors and straw
drift tubes~\cite{LHCb-DP-2013-003,LHCb-DP-2017-001} placed downstream of the magnet.
The tracking system provides a measurement of the momentum, \ptot, of charged particles with
a relative uncertainty that varies from 0.5\% at low momentum to 1.0\% at 200\gevc.
The minimum distance of a track to a primary $pp$ collision vertex (PV), the impact parameter (IP), 
is measured with a resolution of $(15+29/\pt)\mum$,
where \pt is the component of the momentum transverse to the beam, in\,\gevc.
Different types of charged hadrons are distinguished using information
from two ring-imaging Cherenkov detectors~\cite{LHCb-DP-2012-003}. 
Photons, electrons and hadrons are identified by a calorimeter system consisting of
scintillating-pad and preshower detectors, an electromagnetic
and a hadronic calorimeter. Muons are identified by a
system composed of alternating layers of iron and multiwire
proportional chambers~\cite{LHCb-DP-2012-002}.
The online event selection is performed by a trigger~\cite{LHCb-DP-2012-004}, 
which consists of a hardware stage, based on information from the calorimeter and muon
systems, followed by a software stage, which applies a full event
reconstruction.\\
 At the hardware trigger stage, events are required to have a muon with high \pt or a
  hadron, photon or electron with high transverse energy in the calorimeters. 
  The software trigger requires a two-, three- or four-track
  secondary vertex (SV) with a significant displacement from any primary
  $pp$ interaction vertex and a multivariate algorithm~\cite{BBDT,LHCb-PROC-2015-018} is used for
  the identification of SVs consistent with the decay
  of a \bquark hadron.

 Simulation is required to model the effects of the detector acceptance and the selection requirements. Simulated $B^{0} \to p \bar{p} K^{+} \pi^{-}$ decays are generated with an uniform distribution over phase space.  
  In the simulation, $pp$ collisions are generated using
  \pythia~\cite{Sjostrand:2007gs,*Sjostrand:2006za} 
  with a specific \lhcb configuration~\cite{LHCb-PROC-2010-056}.
  Decays of unstable particles
  are described by \evtgen~\cite{Lange:2001uf}, in which final-state
  radiation is generated using \photos~\cite{davidson2015photos}.
  The interaction of the generated particles with the detector, and its response,
  are implemented using the \geant
  toolkit~\cite{Allison:2006ve, *Agostinelli:2002hh} as described in
  Ref.~\cite{LHCb-PROC-2011-006}.

\section{Selection}
The \mbox{$B^{0} \to p\bar{p} K^{+}\pi^{-}$} candidates are formed by combining four charged hadron
candidates: a proton, an antiproton, as well as a kaon and a pion of opposite electric charges. Candidates are selected using a filtering stage followed by a selection based on a boosted decision tree (BDT) classifier~\cite{Breiman1983ClassificationAR} and on particle identification (PID) requirements. 

In the filtering stage, the final-state tracks are selected by requiring $\pt>0.3\gevc$, $p>1.5\gevc$ and the sum of their \pt greater than 1.8\gevc. To ensure that the $B^{0}$ candidate is produced in the primary interaction, a tight requirement on the direction angle, $\theta$, between the reconstructed $B^{0}$ momentum and the distance vector between the associated PV and the $B^{0}$ decay vertex is imposed ($\cos\theta > 0.9999$). Moreover, in order to exclude final-state particles coming directly from the PV, a requirement of $\chisqip>8, 5, 3$ is imposed respectively to
pion, kaon and proton candidates, where $\chisqip$ is defined as the difference between the vertex-fit $\chi^{2}$ of a PV reconstructed with and without the considered track. 
The different $\chisqip$ requirements reflect the different amount of background expected for each particle type.

A BDT classifier is then used to further suppress combinatorial background. The input variables are: the $\chisqip$ and the flight distance of the $B^{0}$ candidate; the quality of the $B^{0}$ vertex; the minimum $p$ and \pt between proton and antiproton; the largest distance of closest approach between any pair of tracks belonging to the signal candidate; and the pointing variable defined as $|\Vec{p}_{B}|\sin\theta/(|\Vec{p}_{B}|\sin\theta + \sum_{i} |\Vec{p}_{i}|\sin\theta_{i})$ 
where $\Vec{p}_{B}$ is the momentum of the $B^0$ candidate, $\Vec{p}_{i}$ is the momentum of daughter $i$ and $\theta_{i}$ is the angle between $\Vec{p}_{i}$ and the vector connecting the primary and secondary vertices. 

The BDT classifier is trained using simulated \mbox{$B^{0} \to p\bar{p} K^{+}\pi^{-}$} decays as signal and candidates in the $B^{0}$ invariant-mass region above the signal, $5450 < m_{p \bar p K^{+} \pi^{-}} < 5550\gevcc$, as background. Tight PID requirements are applied to suppress cross-feed background from other $b$-hadron decays, where one final-state particle is misidentified.
An optimised combination of BDT and PID requirements is chosen in order to maximise the figure of merit $S/\sqrt{S+B}$, where $S (B)$ is the signal (background) yield, giving a signal retention of about $64\% $ and a background rejection of more than $98\%$.
After all selection requirements are applied, 4\% of the events have multiple candidates. For these events, one candidate is chosen randomly.

To reject intermediate charm resonances, candidates with a $K^{+}\pi^{-}$ invariant mass compatible with the $\Dzb$ meson mass and a $\bar{p}K^{+}\pi^{-}$ invariant mass compatible with the $\Lcbar$ baryon mass are removed. In order to exclude charmonium contributions, the $p \bar{p}$ invariant mass is required to be less than 2.85\gevcc.   
The vetoed candidates corresponding to the charmed $B^{0} \to p\bar{p} \Dzb (\to K^{+} \pi^{-})$, which have the same final state as the  signal decay, are retained as a control channel for systematic studies.
 
The $p\bar{p}K^{+}\pi^{-}$ invariant-mass distributions after the selection are shown in Fig.~\ref{fig:Run1_data_fit}. A few sources of background contribute into the considered $B^{0}$ invariant-mass region and consist mainly of $b$-hadron decays where final-state hadrons are not correctly identified. Partially reconstructed decays are also present in the low-mass region, but do not constitute a peaking background. All these background sources are included in the baseline fit model described in Sec.~\ref{Asymmetries}.

\section{Measurement of asymmetries}\label{Asymmetries}
\begin{figure}
    \centering
    \includegraphics[scale=0.39]{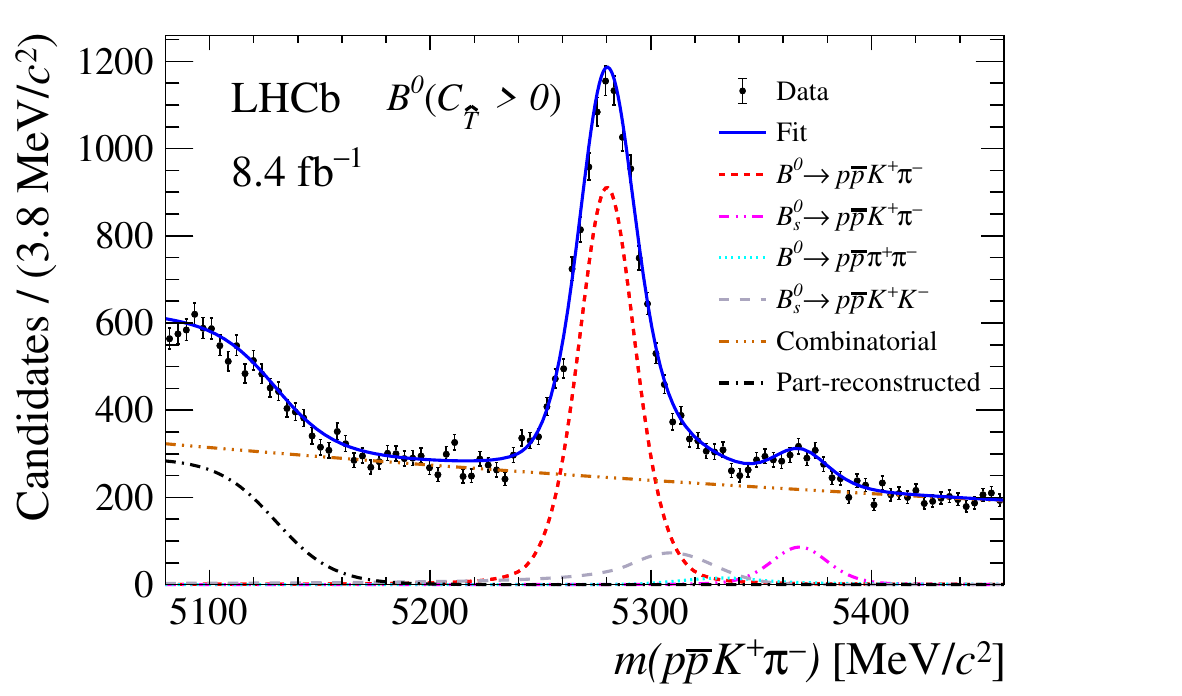}
    \includegraphics[scale=0.39]{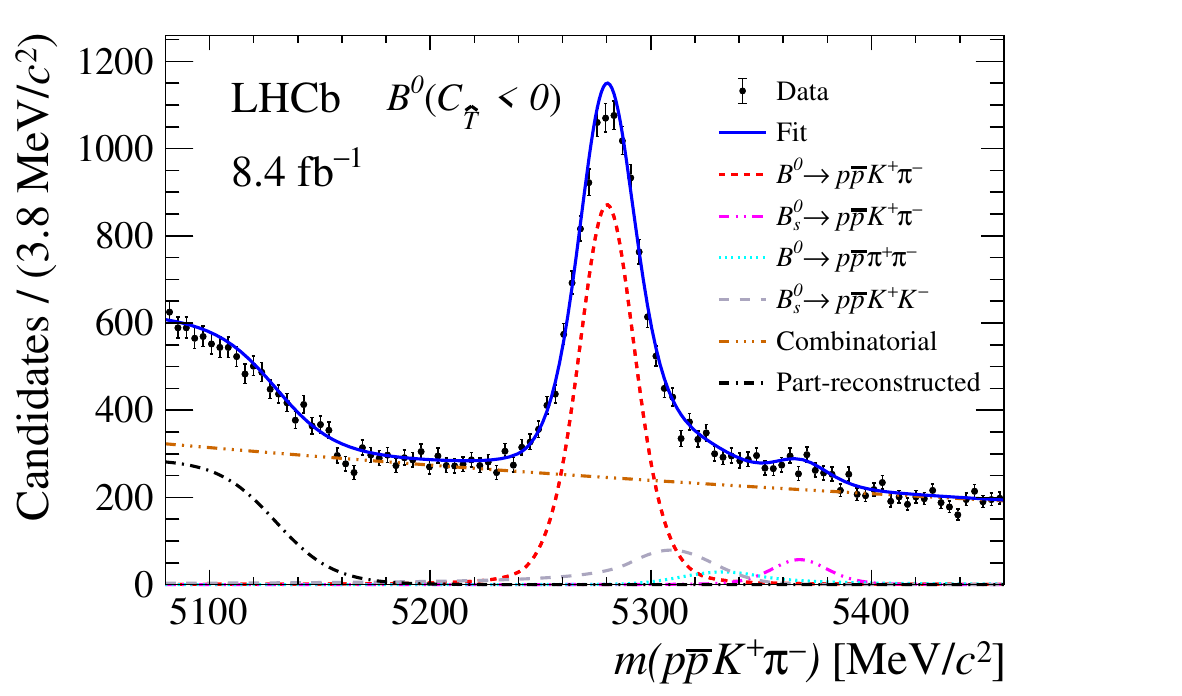}
    \includegraphics[scale=0.39]{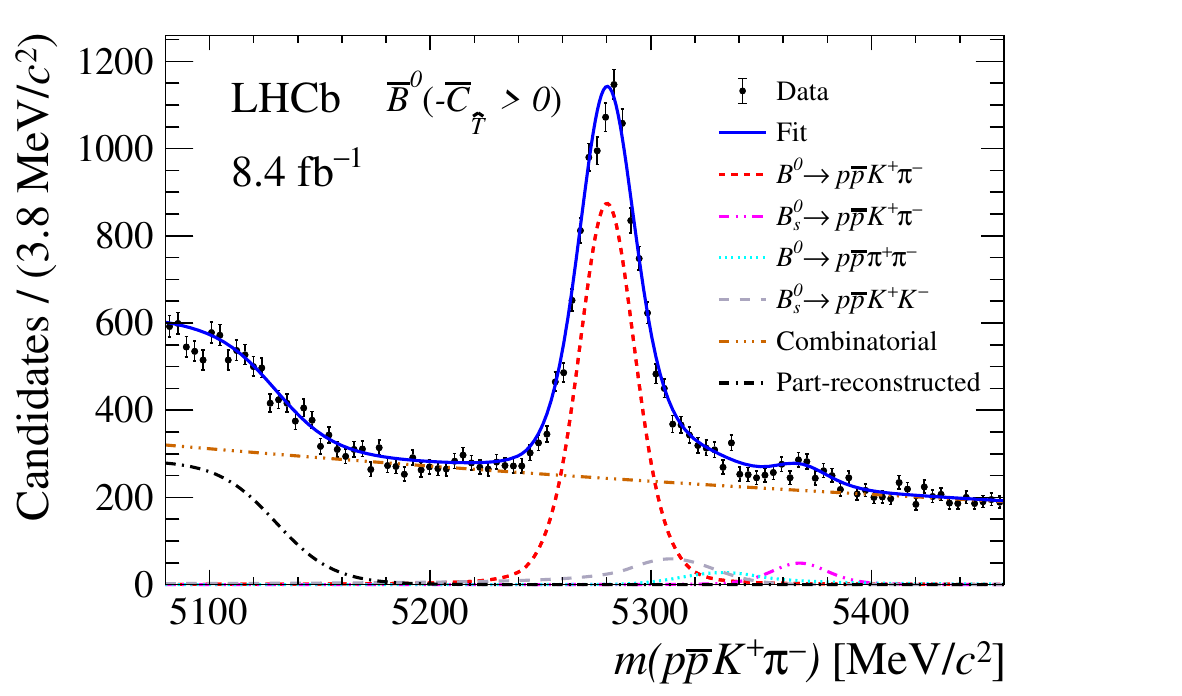}
    \includegraphics[scale=0.39]{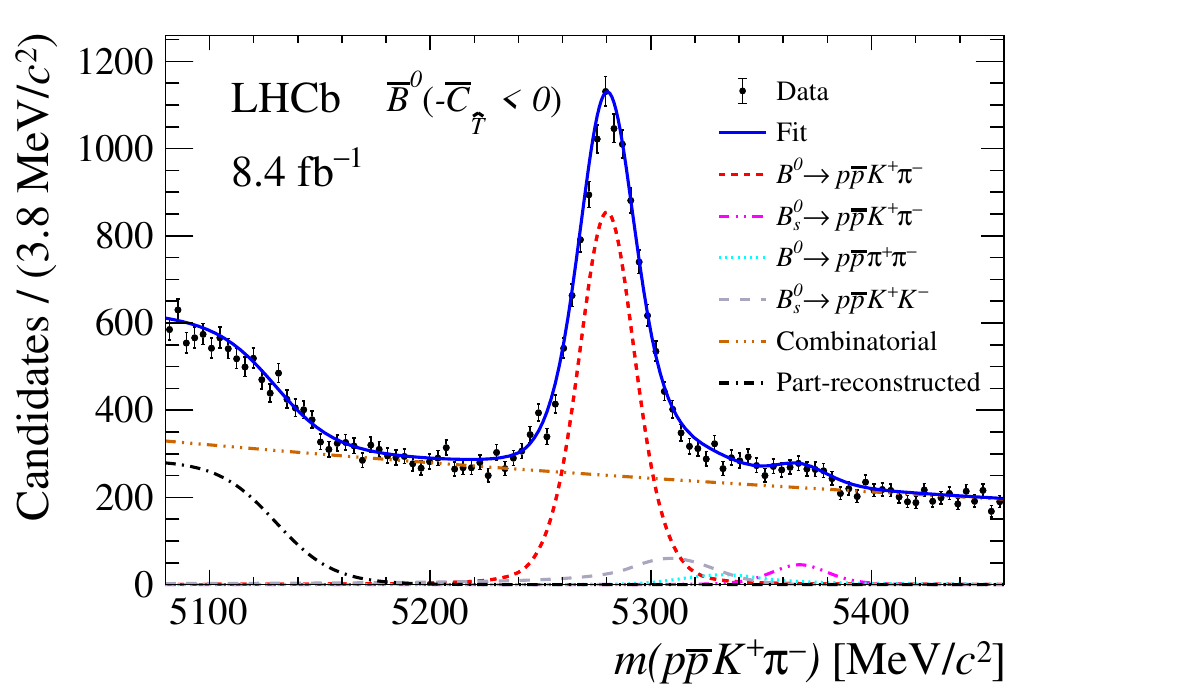}
    \caption{Distributions for combined Run 1 and Run 2 data of the $p\bar{p}K^{+}\pi^{-}$ invariant mass in the four samples defined by $B^{0}$ ($\bar{B}^{0}$) flavour and the sign of $C_{\hat{T}}$ ($\bar{C}_{\hat{T}}$). The results of the fit, as described in the legend, are overlaid on the data.}
    \label{fig:Run1_data_fit}
\end{figure}

The asymmetries for $B^{0}\to p\bar{p}K^{+}\pi^{-}$ decays are measured using an extended maximum likelihood fit to the $m_{p\bar{p}K^{+}\pi^{-}}$ distributions.  
The selected data sample is split into four subsamples according to the $B^{0}$ ($\bar{B}^{0}$) flavour and the sign of $C_{\hat{T}}$ ($\bar{C}_{\hat{T}}$). A simultaneous fit to the $m_{p\bar p K^{+}\pi^{-}}$ distributions of
the four subsamples in Fig.~\ref{fig:Run1_data_fit} is used to determine the number of signal and background yields and
the asymmetries $A_{\hat{T}}$ and $\bar{A}_{\hat{T}}$.
The two asymmetries $A_{T}$ and $\bar A_{T}$ are
included in the fit model as 
\begin{equation}
N_{B^{0},C_{\hat{T}}>0} = \frac{1}{2}N_{B^{0}}(1+A_{\hat{T}}),
\end{equation}
\begin{equation}
N_{B^{0},C_{\hat{T}}<0} = \frac{1}{2}N_{B^{0}}(1-A_{\hat{T}}),
\end{equation}
\begin{equation}
N_{\bar B^{0},-\bar{C}_{\hat{T}}>0} = \frac{1}{2}N_{\bar
  B^{0}}(1+\bar{A}_{\hat{T}}),
\end{equation}
\begin{equation}
N_{\bar B^{0},-\bar{C}_{\hat{T}}<0} = \frac{1}{2}N_{\bar B^{0}}(1-\bar{A}_{\hat{T}}),
\end{equation}
where $N$ denotes the number of $B^{0}$ and $\bar B^{0}$ satisfying the requirement on $C_{\hat T}$ and $\bar{C}_{\hat T}$. The $P$- and \CP-violating asymmetries, $a^{\hat{T}\text{-odd}}_{P}$  and $a^{\hat{T}\text{-odd}}_{\CP}$, are then obtained according to Eq.~\ref{eq:atodd}. The correlations between the $A_{\hat{T}}$ and $\bar A_{\hat{T}}$ asymmetries are verified to be negligible.

The
invariant-mass distributions of the $B^{0}\to p\bar{p}K^{+}\pi^{-}$ and $B^{0}_{s} \to p\bar{p}K^{+}\pi^{-}$ decays are both modelled by a Hypatia function~\cite{MartinezSantos:2013ltf}, with mean and common width determined from
data. All other fit parameters for these decays are taken from simulation. 
The two main sources of cross-feed backgrounds are due to $B^{0}\to p\bar{p}K^{+}K^{-}$ and $B^{0}\to p\bar{p}\pi^{+}\pi^{-}$ decays, where a kaon or a pion is misidentified. They are modelled with a double Crystal Ball function~\cite{Skwarnicki:1986xj} with all shape parameters fixed according to simulation. The combinatorial background is parameterised with an exponential
function where the parameters are left free to vary in the fits. Partially reconstructed background is described by a function of the form 
$ f(m) = (e^{c(m - m_{0})}+1)^{-1}$,
where the parameters $m_{0}$ and $c$ are determined from data.

Two different approaches are followed to search for $P$ and \CP violation: a measurement integrated over the phase space, with the charmonium region removed, and measurements in different regions of the phase space. In multi-body decays, \CP asymmetries may vary over the phase space due to resonant contributions and their interference effects, possibly canceling when integrated over the whole phase space. To enhance the sensitivity to \CP violation, measurements in different regions of the phase space are performed. No $P$-odd amplitude information from an amplitude analysis is available for the $B^{0} \to p \bar p K^{+}\pi^{-}$ decay to provide information on the most interesting phase-space regions.
The results of the first approach are obtained by fitting the phase-space integrated data sample divided according to the $B$ flavour and the sign of the triple product. The fit result is shown in Fig.~\ref{fig:Run1_data_fit}. 

The measurements in different regions of phase space are performed by dividing the sample using a binning scheme based on the invariant masses of the $K^{+}\pi^{-}$ and $p
  \bar p$ combinations, $m_{K^{+}\pi^{-}}$ and $m_{p
  \bar p}$, the cosine of the angle of
the $K^{+}$ ($p$) with respect to the opposite direction to the $B^{0}$ momentum in the $K^{+}\pi^{-}$ ($p\bar p$) rest frame, $\cos\theta_{K^{+}\pi^{-}}$ ($\cos\theta_{p
  \bar p}$), and the angle between the planes defined by the
$K^{+}\pi^{-}$ and $p\bar p$ tracks in the $B^{0}$ rest frame, $\phi$. 
The background-subtracted distributions of $m_{K^{+}\pi^{-}}$, $m_{p
  \bar p}$, $\cos\theta_{K^{+}\pi^{-}}$, $\cos\theta_{p
  \bar p}$ and $\phi$ for $B^{0}$ ($\bar{B}^{0}$) candidates with $C_{\hat{T}} > 0$ and $C_{\hat{T}} < 0$ 
($-\bar{C}_{\hat{T}}>0$ and $-\bar{C}_{\hat{T}}<0$) are shown in Fig.~\ref{fig:background_subracted_phs_Run1} of Appendix~\ref{sec:bkgsubtr}.

Two different schemes, chosen before examining the data to avoid possible biases, are used to divide the phase space. 
The phase space is divided into 24 (40) regions and the definition of the scheme A (B) is reported in Table~\ref{tab:schemeconf1} (Table~\ref{tab:schemeconf2}) of Appendix~\ref{binning_scheme}.
In binning scheme B, some region edges in the $m(K^{+}\pi^{-})$ variable correspond to the resonance mass pole where the strong phase changes sign. This choice further enhances the sensitivity to \CP violation. Due to many overlapping resonances in the $K^{+}\pi^{-}$ mass spectrum, only the $K^{*}(892)^{0}$ and $K_{2}^{*}(1430)$ states, for which the peaks can be clearly identified, are split.

\begin{figure}
    \centering
    \includegraphics[scale=0.39]{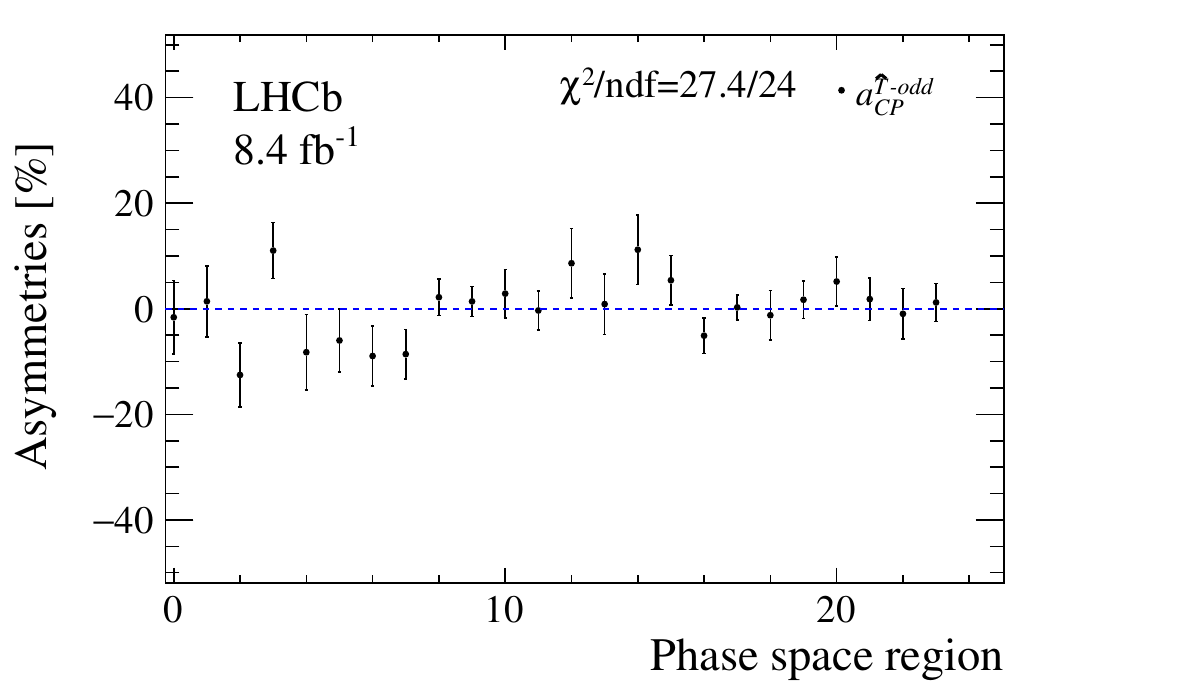}
    \includegraphics[scale=0.39]{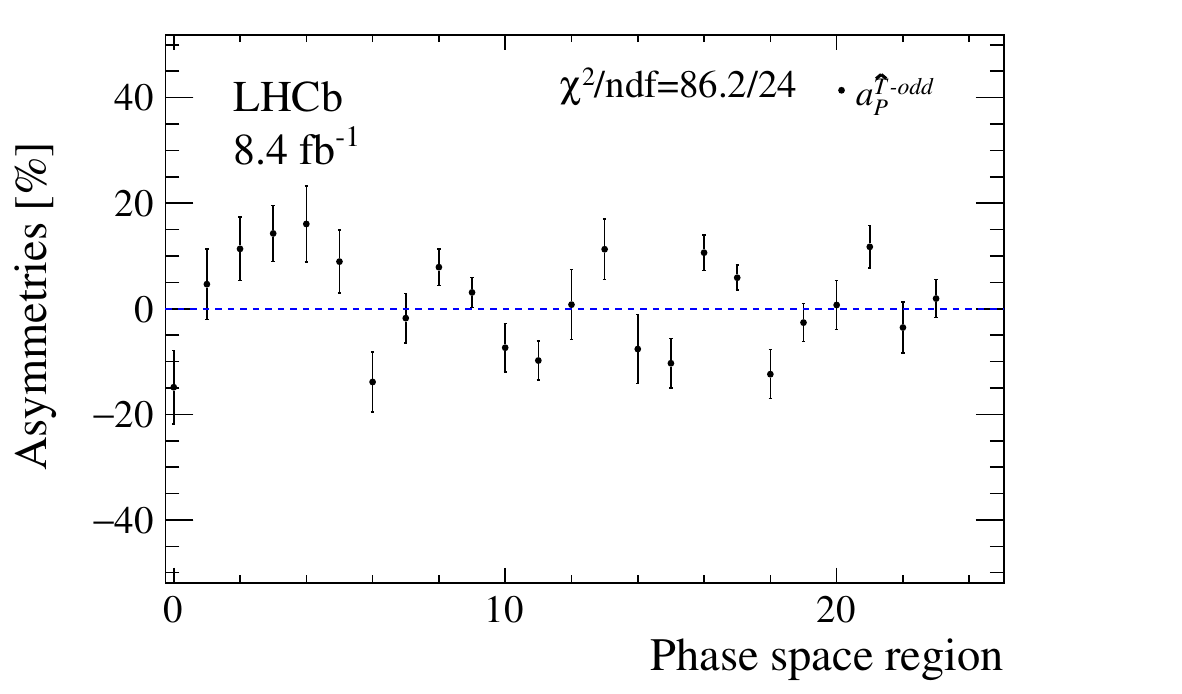}
    \caption{The $a^{\hat{T}\text{-odd}}_{\CP}$ (left) and $a^{\hat{T}\text{-odd}}_{P}$ (right) asymmetry parameters in each region of the phase space for Run 1 and Run 2 data combined for binning scheme A. The error bars represent the sum in quadrature of the statistical and systematic uncertainties. The $\chi^{2}$ per number of degrees of freedom (ndf) is calculated with respect to the null hypothesis. }
    \label{fig:acp_run12_full}
\end{figure}

\begin{figure}
    \centering
    \includegraphics[scale=0.39]{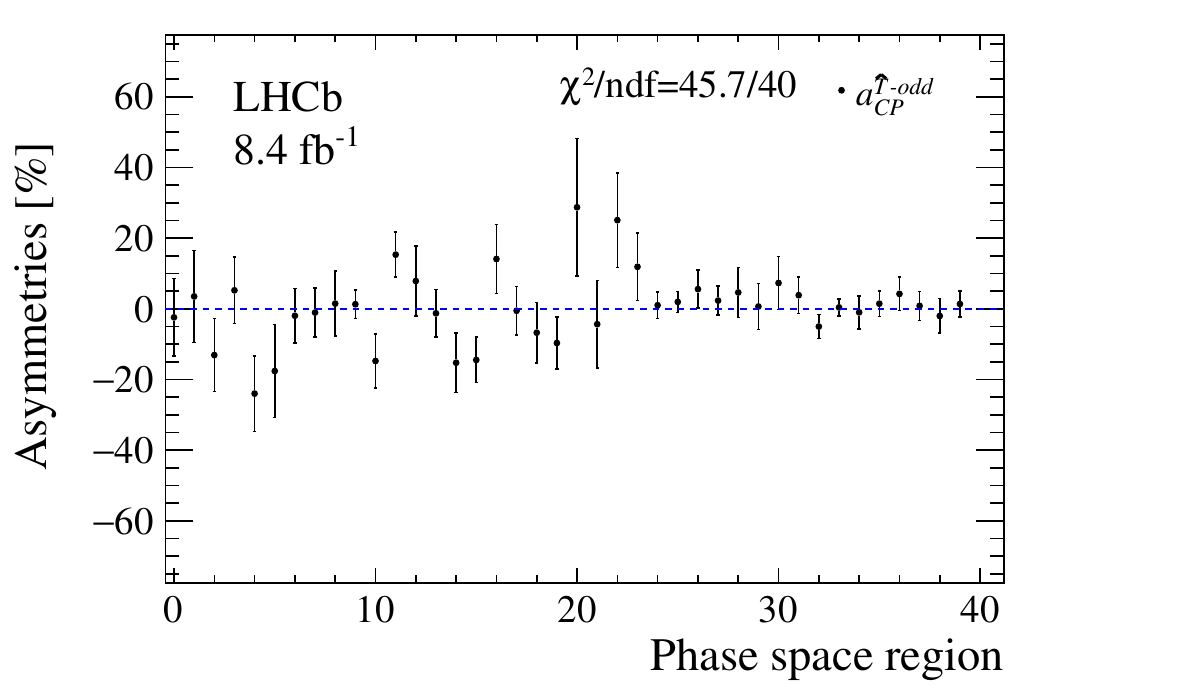}
    \includegraphics[scale=0.39]{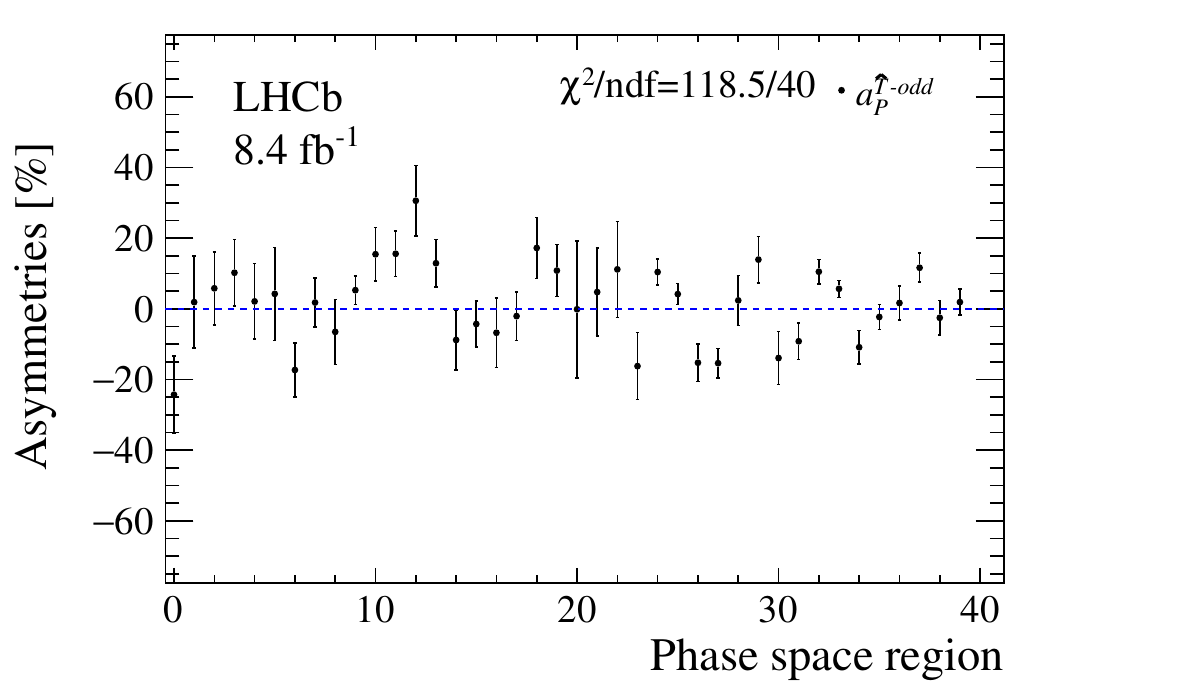}
    \caption{The $a^{\hat{T}\text{-odd}}_{\CP}$ (left) and $a^{\hat{T}\text{-odd}}_{P}$ (right) asymmetry parameters in each region of the phase space for Run 1 and Run 2 data combined for binning scheme B. The error bars represent the sum in quadrature of the statistical and systematic uncertainties. The $\chi^{2}$ per ndf is calculated with respect to the null hypothesis.}
    \label{fig:acp_run12_fullB}
\end{figure}

The same fit model used for the integrated measurement is exploited to fit the $B^{0}$ mass distribution separately for each phase-space region. The distributions of the measured asymmetries for scheme A (B) are shown in Fig.~\ref{fig:acp_run12_full} (Fig.~\ref{fig:acp_run12_fullB}) and the numerical results are reported in Table~\ref{tab:measured_asymm_combined}~(Table~\ref{tab:measured_asymm_combinedB}) in Appendix~\ref{binning_scheme}.


The compatibility with the \CP ($P$) conservation hypothesis is tested by means of a $\chi^{2}$ test, where the $\chi^{2}$ is defined as $X^{T}V^{-1}X$, with $X$ denoting the array of $a^{\hat{T}\text{-odd}}_{\CP}$ ($a^{\hat{T}\text{-odd}}_{P}$) measurements, $V^{-1}$ is the inverse of the covariance matrix $V$, defined as the sum of the statistical and systematic covariance matrices. 
An average systematic uncertainty, whose evaluation is discussed in Sec.~\ref{systematics}, is assumed for each bin. The statistical and systematic uncertainties are considered uncorrelated among the bins. No  
significant $\CP$ violation is observed with either of the binning schemes, while some phase-space regions exhibit $P$-violation.

\section{Systematic uncertainties and cross-checks}\label{systematics}
The systematic uncertainties are determined in each phase space region with simulated pseudoexperiments having the same number of signal candidates as the real data and the largest value found in a single bin is used as the systematic error representative for all bins.
The sources of systematic uncertainty and their relative contributions, expressed as a percentage of the statistical uncertainty, are listed in Table~\ref{tab:systematics_table}. The contributions are uncorrelated and thus added in quadrature.
The systematic uncertainty related to the detector resolution, which could introduce a migration of signal decays between $C_{\hat{T}} > 0$ and $C_{\hat{T}} < 0$ ($-\bar{C}_{\hat{T}}>0$ and $-\bar{C}_{\hat{T}}<0$) categories for $B^{0}$ ($\bar{B}^{0}$), is estimated in every region of the phase space using a simulated
sample of $B^{0} \to p\bar{p}K^{+} \pi^{-}$ decays.
The difference between the reconstructed and generated asymmetry is considered as systematic uncertainty. Although the difference is negligibly small, a common relative uncertainty of 1\% is assigned to every phase space region.
To test whether the fit procedure introduces any bias, pseudoexperiments are generated
from the baseline fit model using the measured asymmetry values and fitting them with the same
model. Since the observed bias is compatible with zero, the largest value among the mean of the pull distribution estimated in every single region of the phase space, 5\%, is assigned as the fit procedure relative systematic uncertainty. 
The systematic uncertainties related to the choice of the model for the signal and background components are evaluated by using alternative models that have comparable fit quality with the baseline model. A double Crystal Ball function is used for the signal while the background is described by a linear function. 
Pseudoexperiments are generated using the alternative model and the baseline fit model is then used to fit each generated sample. 
Since the observed bias is not significantly different from zero, the largest value among the mean of the pull distribution estimated in every single region of the phase space, 5\%, is assigned as alternative fit model systematic uncertainty. 
The effect of fixing the mass and resolutions of the cross-feed
background from simulated samples is assessed by varying their values. The mass and the resolution are varied uniformly in a $\pm 3\mevcc$ and $\pm 5\mevcc$ range respectively around the values found in the simulation. The range  
is chosen in order to take into account the possible discrepancy between simulation and data for the $B^{0}$ mass and resolution. A relative contribution of 5\%, which corresponds to the largest value among the mean of the pull distribution estimated in every single region of the phase space, is assigned as systematic uncertainty. 
As a cross-check, a possible experimental bias is tested by measuring the $a^{\hat{T}\text{-odd}}_{\CP}$ asymmetry using the \mbox{$B^{0} \to p\bar{p} \Dzb (\to K^{+} \pi^{-})$} control channel. Since negligible \CP violation is expected for this channel, which proceeds through the tree-level $b \to cu\bar{u}$ transition, any deviation of the \CP asymmetry
from zero is considered as a bias introduced by the experimental reconstruction and analysis technique. The asymmetry measured on the \mbox{$B^{0} \to p\bar{p} \Dzb (\to K^{+} \pi^{-})$} control sample, $a^{\hat{T}\text{-odd}}_{\CP}=(-1.0 \pm 1.5)\%$, shows no significant bias.
The test is repeated for different
regions of phase space, using a control sample weighted according to the kinematic distributions of the signal and for different magnet polarities, and gives consistent results. Further cross-checks are made to test the stability of the results with respect to the different magnet polarities, the choice made in the selection of multiple candidates, and the effect of the trigger and selection criteria. No systematic uncertainty is assigned since all these checks give results compatible with the nominal ones. In addition, correlations between $C_{\hat{T}}$ and ${\bar C}_{\hat{T}}$ and the kinematic and topological variables used in the selection are checked using both simulated samples and background-subtracted data and found to be negligible. 

\begin{table}
\caption{Sources of systematic uncertainty and their relative contributions expressed as a percentage of the statistical uncertainty. In order to obtain the absolute systematic uncertainty assigned to specific region of the phase space, the numbers reported here have to be multiplied by the corresponding statistical uncertainties. $\Delta a^{\hat{T}\text{-odd}}_{\CP}$ and $\Delta a^{\hat{T}\text{-odd}}_{P}$ indicate the uncertainty assigned to $a^{\hat{T}\text{-odd}}_{\CP}$ and $a^{\hat{T}\text{-odd}}_{P}$ respectively.}
\begin{center}
\begin{tabular}{ l S S} 
\hline
Contribution & $\Delta a^{\hat{T}\text{-odd}}_{\CP}$ [\%]& $\Delta a^{\hat{T}\text{-odd}}_{P}$ [\%]\\
\hline
Detector resolution & 1 &  1\\ 
Fit procedure & 5 &  5\\ 
Alternative fit & 5 &  5\\ 
Mass resolution & 5 & 5 \\ 
\hline
Total & 9 & 9 \\
\hline 
\end{tabular}
\end{center}
\label{tab:systematics_table}
\end{table}

\section{Results and conclusion}
In conclusion, a search for $P$ and \CP violation in $B^{0}\to p\bar{p}K^{+}\pi^{-}$ decays is performed both globally and in regions of the phase space.
The measured phase-space integrated asymmetries are
\begin{equation*}
  a^{\hat{T}\text{-odd}}_{P} = (1.49\pm 0.85 \pm 0.08)\%,
\end{equation*}
\begin{equation*}
  a^{\hat{T}\text{-odd}}_{\CP} = (0.51\pm 0.85 \pm 0.08)\%,
\end{equation*}
where the uncertainties are respectively statistical and systematic.
Both are consistent with $P$ and \CP conservation. The 1\% theoretical prediction discussed in Sec.~\ref{sec:IntroductionB2ppbarKpi} is beyond the reach of the current experimental sensitivity. However, an observation of \CP violation in the $B^{0}\to p\bar{p}K^{+}\pi^{-}$ decay would have been a hint of Beyond Standard Model contributions.
Measurements in regions of the phase space are consistent with the \CP-symmetry hypothesis with a $p$-value of 0.28 (0.24), according to $\chi^{2}=27.4/24$ (\mbox{$\chi^{2}=45.7/40$}), corresponding to $1.1\,\sigma$ ($1.2\,\sigma$) deviation for scheme A (scheme B).   For $P$-symmetry, a $p$-value of 6.1$\times 10^{-9}$ (1.1$\times 10^{-9}$) is found, according to $\chi^{2}=86.2/24$ (\mbox{$\chi^{2}=118.5/40$}), corresponding to $5.8\sigma$ ($6.0\sigma$) deviation for scheme A (scheme B). Significant $P$-asymmetries are observed in the region of low $p\bar{p}$ mass and near the $K^{*}(892)^{0}$ resonance. However, a full amplitude analysis of the $B^{0} \to p \bar{p}K^{+}\pi^{-}$ decay is needed to associate the observed $P$-parity violation with any underlying resonance amplitude.
In conclusion, the data are consistent with $P$-parity violation, but show no evidence for \CP violation. 
 
\section*{Acknowledgements}
\noindent We express our gratitude to our colleagues in the CERN
accelerator departments for the excellent performance of the LHC. We
thank the technical and administrative staff at the LHCb
institutes.
We acknowledge support from CERN and from the national agencies:
CAPES, CNPq, FAPERJ and FINEP (Brazil); 
MOST and NSFC (China); 
CNRS/IN2P3 (France); 
BMBF, DFG and MPG (Germany); 
INFN (Italy); 
NWO (Netherlands); 
MNiSW and NCN (Poland); 
MEN/IFA (Romania); 
MICINN (Spain); 
SNSF and SER (Switzerland); 
NASU (Ukraine); 
STFC (United Kingdom); 
DOE NP and NSF (USA).
We acknowledge the computing resources that are provided by CERN, IN2P3
(France), KIT and DESY (Germany), INFN (Italy), SURF (Netherlands),
PIC (Spain), GridPP (United Kingdom), 
CSCS (Switzerland), IFIN-HH (Romania), CBPF (Brazil),
Polish WLCG  (Poland) and NERSC (USA).
We are indebted to the communities behind the multiple open-source
software packages on which we depend.
Individual groups or members have received support from
ARC and ARDC (Australia);
Minciencias (Colombia);
AvH Foundation (Germany);
EPLANET, Marie Sk\l{}odowska-Curie Actions and ERC (European Union);
A*MIDEX, ANR, IPhU and Labex P2IO, and R\'{e}gion Auvergne-Rh\^{o}ne-Alpes (France);
Key Research Program of Frontier Sciences of CAS, CAS PIFI, CAS CCEPP, 
Fundamental Research Funds for the Central Universities, 
and Sci. \& Tech. Program of Guangzhou (China);
GVA, XuntaGal, GENCAT and Prog.~Atracci\'on Talento, CM (Spain);
SRC (Sweden);
the Leverhulme Trust, the Royal Society
 and UKRI (United Kingdom).

\section*{Appendices}

\appendix

\section{Measured asymmetries in regions of phase space}\label{binning_scheme}
The definitions of the regions of phase space of the four-body $B^{0}\to p\bar{p}K^{+}\pi^{-}$ decay of the two binning schemes are reported in Tables \ref{tab:schemeconf1} and \ref{tab:schemeconf2}. The corresponding measured asymmetries in each region of phase space are reported in Tables \ref{tab:measured_asymm_combined} and \ref{tab:measured_asymm_combinedB}.

\begin{table}[H]
\caption{Definition of the 24 regions that form scheme A for the $B^{0}\to p\bar{p}K^{+}\pi^{-}$ decay.}
\centering
\begin{tabular}{c c c r@{, }l r@{, }l c}
Region & $m_{p\bar{p}}$(MeV/$c^{2}$) & $m_{K^{+}\pi^{-}}$(MeV/$c^{2}$) & \multicolumn{2}{c}{$\cos\theta_{p\bar{p}}$} & \multicolumn{2}{c}{$\cos\theta_{K^{+}\pi^{-}}$} & $\phi$\\
\hline
0 & $(1800,2850)$ & $(500,1200)$ & $(-1$ & $0)$ & $(-1$ &$0)$ & $(0,\pi/2)$ \\
1 & $ (1800,2850)$ & $ (500,1200)$ & $ (-1$ & $0)$ & $(-1$ & $0)$ & $ (\pi/2,\pi)$ \\
2 & $(1800,2850)$ & $(500,1200)$ & $ (-1$ & $0)$ & $(0$ & $1)$ & $(0, \pi/2)$ \\
3 & $ (1800,2850)$ & $ (500,1200)$ & $ (-1$ & $0)$ & $ (0$ & $1)$ & $(\pi/2, \pi)$ \\
4 & $ (1800,2850)$ & $(500,1200)$ & $ (0$ & $1)$ & $ (-1$ & $0)$ & $ (0, \pi/2)$ \\
5 & $ (1800,2850)$ & $(500,1200)$ & $ (0$ & $1)$ & $ (-1$ & $0)$ & $(\pi/2, \pi)$ \\
6 & $ (1800,2850)$ & $ (500,1200)$ & $ (0$ & $1)$ & $ (0$ & $1)$ & $ (0, \pi/2)$ \\
7 & $ (1800,2850)$ & $ (500,1200)$ & $  (0$ & $1)$ & $ (0$ & $1)$ & $ (\pi/2, \pi)$ \\
8 & $ (1800,2850)$ & $ (1200,2200)$ & $ (-1$ & $0)$ & $ (-1$ & $0)$ & $(0, \pi/2)$ \\
9 & $(1800,2850)$ & $ (1200,2200)$ & $ (-1$ & $0)$ & $ (-1$ & $0)$ & $ (\pi/2, \pi)$ \\
10 & $ (1800,2850)$ & $(1200,2200)$ & $ (-1$ & $0)$ & $ (0$ & $1)$ & $ (0, \pi/2)$ \\
11 & $ (1800,2850)$ & $ (1200,2200)$ & $ (-1$ & $0)$ & $ (0$ & $1)$ & $ (\pi/2,\pi)$ \\
12 & $ (1800,2850)$ & $ (1200,2200)$ & $ (0$ & $1)$ & $ (-1$ & $0)$ & $(0, \pi/2)$ \\
13 & $(1800,2850)$ & $ (1200,2200)$ & $ (0$ & $1)$ & $ (-1$ & $0)$ & $(\pi/2, \pi)$ \\
14 & $(1800,2850)$ & $ (1200,2200)$ & $ (0$ & $1)$ & $(0$ & $1)$ & $ (0, \pi/2)$ \\
15 & $ (1800,2850)$ & $(1200,2200)$ & $ (0$ & $1)$ & $ (0$ & $1)$ & $ (\pi/2, \pi)$ \\
16 & $ (1800,2850)$ & $ (2200,3600)$ & $ (-1$ & $0)$ & $ (-1$ & $0)$ & $ (0, \pi/2)$ \\
17 & $(1800,2850)$ & $(2200,3600)$ & $ (-1$ & $0)$ & $ (-1$ & $0)$ & $(\pi/2, \pi)$ \\
18 & $ (1800,2850)$ & $ (2200,3600)$ & $(-1$ & $0)$ & $ (0$ & $1)$ & $ (0, \pi/2)$ \\
19 & $(1800,2850)$ & $(2200,3600)$ & $ (-1$ & $0)$ & $ (0$ & $1)$ & $ (\pi/2, \pi)$ \\
20 & $ (1800,2850)$ & $ (2200,3600)$ & $ (0$ & $1)$ & $ (-1$ & $0)$ & $ (0, \pi/2)$ \\
21 & $(1800,2850)$ & $(2200,3600)$ & $ (0$ & $1)$ & $ (-1$ & $0)$ & $ (\pi/2, \pi)$ \\
22 & $(1800,2850)$ & $ (2200,3600)$ & $ (0$ & $1)$ & $ (0$ & $1)$ & $(0, \pi/2)$ \\
23 & $(1800,2850)$ & $ (2200,3600)$ & $(0$ & $1)$ & $ (0$ & $1)$ & $ (\pi/2, \pi)$ \\
\hline
\end{tabular}
\label{tab:schemeconf1}
\end{table}

\begin{table}[H]
\caption{Measurements of $a^{\hat{T}\text{-odd}}_{CP}$ and $a^{\hat{T}\text{-odd}}_{P}$ in specific phase-space regions for the
 $B^{0}\to p\bar{p} K^{+}\pi^{-}$ decay for binning scheme A. Each value is obtained through an independent fit to the $B^{0}$ invariant-mass distribution of the candidates in the corresponding region of the phase space. The uncertainties are only statistical.}
\centering
\begin{tabular}{c r@{\,$\pm$\,}l r@{\,$\pm$\,}l r@{\,$\pm$\,}l r@{\,$\pm$\,}l}
Region & \multicolumn{2}{c}{$A_{\hat{T}}(\%)$} & \multicolumn{2}{c}{$\bar{A}_{\hat{T}}(\%)$} & \multicolumn{2}{c}{$a^{\hat{T}\text{-odd}}_{CP}(\%)$} & \multicolumn{2}{c}{$a^{\hat{T}\text{-odd}}_{P}(\%)$} \\\hline
0 & $-16.5$ & 10.1 & $-13.2$ & 9.5 & $-1.6$&7.0 & $-14.9$&7.0 \\1 & 6.1&9.2 & 3.2&9.8 & 1.4&6.7 & 4.7&6.7 \\2 & $-1.2$&7.0 & 23.9&10.0 & $-12.5$&6.0 & 11.4&6.0 \\3 & 25.3&7.2 & 3.2&7.8 & 11.0&5.3 & 14.3&5.3 \\4 & 7.8&11.1 & 24.3&9.0 & $-8.2$&7.1 & 16.1&7.1 \\5 & 2.9&8.3 & 14.9&8.6 & $-6.0$&6.0 & 8.9&5.9 \\6 & $-22.8$&7.4 & $-4.9$&8.6 & $-8.9$&5.7 & $-13.9$&5.7 \\7 & $-10.4$&6.8 & 6.8&6.6 & $-8.6$&4.7 & $-1.8$&4.7 \\8 & 10.1&5.0 & 5.7&4.9 & 2.2&3.5 & 7.9&3.5 \\9 & 4.5&4.0 & 1.7&4.0 & 1.4&2.8 & 3.1&2.8 \\10 & $-4.5$&6.5 & $-10.2$&6.5 & 2.9&4.6 & $-7.4$&4.6 \\11 & $-10.1$&5.2 & $-9.5$&5.2 & $-0.3$&3.7 & $-9.8$&3.7 \\12 & 9.4&9.2 & $-7.8$&9.5 & 8.6&6.6 & 0.8&6.6\\13 & 12.2&8.2 & 10.4&8.0 & 0.9&5.7 & 11.3&5.7 \\14 & 3.6&9.8 & $-18.8$&8.7 & 11.2&6.6 & $-7.6$&6.6 \\15 & $-4.9$&6.0 & $-15.7$&7.1 & 5.4&4.7 & $-10.3$&4.7 \\16 & 5.5&4.8 & 15.7&4.7 & $-5.1$&3.4 & 10.6&3.4 \\17 & 6.2&3.4 & 5.6&3.3 & 0.3&2.4 & 5.9&2.4 \\18 & $-13.6$&6.7 & $-11.2$&6.4 & $-1.2$&4.6 & $-12.4$&4.6 \\19 & $-0.9$&5.1 & $-4.3$&5.1 & 1.7&3.6 & $-2.6$&3.6 \\20 & 5.9&6.2 & $-4.4$&6.9 & 5.2&4.7 & 0.7&4.7 \\21 & 13.6&5.6 & 9.9&5.8 & 1.8&4.0 & 11.7&4.0 \\22 & $-4.5$&6.9 & $-2.6$&6.6 & $-0.9$&4.8 & $-3.5$&4.8 \\23 & 3.1&5.0 & 0.7&5.2 & 1.2&3.6 & 1.9 & 3.6 \\\hline
\end{tabular}

\label{tab:measured_asymm_combined}
\end{table}

\begin{table}
\caption{Definition of the 40 regions that form scheme B for the $B^{0}\to p\bar{p}K^{+}\pi^{-}$ decay.}
\centering
\begin{tabular}{c c c r@{, }l r@{, }l c}
Region & $m_{p\bar{p}}$(MeV/$c^{2}$) & $m_{K^{+}\pi^{-}}$(MeV/$c^{2}$) & \multicolumn{2}{c}{$\cos\theta_{p\bar{p}}$} & \multicolumn{2}{c}{$\cos\theta_{K^+\pi^-}$} & $\phi$ \\
\hline
0 & $(1800,2850)$ & $(500,892)$ & $(-1$ & $0)$ & $(-1$ & $0)$ & $(0,\pi/2)$ \\
1 & $(1800,2850)$ & $(500,892)$ & $(-1$ & $0)$ & $(-1$ & $0)$ & $(\pi/2,\pi)$ \\
2 & $(1800,2850)$ & $(500,892)$ & $(-1$ & $0)$ & $(0$ & $1)$ & $(0,\pi/2)$ \\
3 & $(1800,2850)$ & $(500,892)$ & $(-1$ & $0)$ & $(0$ & $1)$ & $(\pi/2,\pi)$ \\
4 & $(1800,2850)$ & $(500,892)$ & $(0$ & $1)$ & $(-1$ & $0)$ & $(0,\pi/2)$ \\
5 & $(1800,2850)$ & $(500,892)$ & $(0$ & $1)$ & $(-1$ & $0)$ & $(\pi/2,\pi)$ \\
6 & $(1800,2850)$ & $(500,892)$ & $(0$ & $1)$ & $(0$ & $1)$ & $(0,\pi/2)$ \\
7 & $(1800,2850)$ & $(500,892)$ & $(0$ & $1)$ & $(0$ & $1)$ & $(\pi/2,\pi)$ \\
8 & $(1800,2850)$ & $(892,1200)$ & $(-1$ & $0)$ & $(-1$ & $0)$ & $(0,\pi/2)$ \\
9 & $(1800,2850)$ & $(892,1200)$ & $(-1$ & $0)$ & $(-1$ & $0)$ & $(\pi/2,\pi)$ \\
10 & $(1800,2850)$ & $(892,1200)$ & $(-1$ & $0)$ & $(0$ & $1)$ & $(0,\pi/2)$ \\
11 & $(1800,2850)$ & $(892,1200)$ & $(-1$ & $0)$ & $(0$ & $1)$ & $(\pi/2,\pi)$ \\
12 & $(1800,2850)$ & $(892,1200)$ & $(0$ & $1)$ & $(-1$ & $0)$ & $(0,\pi/2)$ \\
13 & $(1800,2850)$ & $(892,1200)$ & $(0$ & $1)$ & $(-1$ & $0)$ & $(\pi/2,\pi)$ \\
14 & $(1800,2850)$ & $(892,1200)$ & $(0$ & $1)$ & $(0$ & $1)$ & $(0,\pi/2)$ \\
15 & $(1800,2850)$ & $(892,1200)$ & $(0$ & $1)$ & $(0$ & $1)$ & $(\pi/2,\pi)$ \\
16 & $(1800,2850)$ & $(1200,1430)$ & $(-1$ & $0)$ & $(-1$ & $0)$ & $(0,\pi/2)$ \\
17 & $(1800,2850)$ & $(1200,1430)$ & $(-1$ & $0)$ & $(-1$ & $0)$ & $(\pi/2,\pi)$ \\
18 & $(1800,2850)$ & $(1200,1430)$ & $(-1$ & $0)$ & $(0$ & $1)$ & $(0,\pi/2)$ \\
19 & $(1800,2850)$ & $(1200,1430)$ & $(-1$ & $0)$ & $(0$ & $1)$ & $(\pi/2,\pi)$ \\
20 & $(1800,2850)$ & $(1200,1430)$ & $(0$ & $1)$ & $(-1$ & $0)$ & $(0,\pi/2)$ \\
21 & $(1800,2850)$ & $(1200,1430)$ & $(0$ & $1)$ & $(-1$ & $0)$ & $(\pi/2,\pi)$ \\
22 & $(1800,2850)$ & $(1200,1430)$ & $(0$ & $1)$ & $(0$ & $1)$ & $(0,\pi/2)$ \\
23 & $(1800,2850)$ & $(1200,1430)$ & $(0$ & $1)$ & $(0$ & $1)$ & $(\pi/2,\pi)$ \\
24 & $(1800,2850)$ & $(1430,2200)$ & $(-1$ & $0)$ & $(-1$ & $0)$ & $(0,\pi/2)$ \\
25 & $(1800,2850)$ & $(1430,2200)$ & $(-1$ & $0)$ & $(-1$ & $0)$ & $(\pi/2,\pi)$ \\
26 & $(1800,2850)$ & $(1430,2200)$ & $(-1$ & $0)$ & $(0$ & $1)$ & $(0,\pi/2)$ \\
27 & $(1800,2850)$ & $(1430,2200)$ & $(-1$ & $0)$ & $(0$ & $1)$ & $(\pi/2,\pi)$ \\
28 & $(1800,2850)$ & $(1430,2200)$ & $(0$ & $1)$ & $(-1$ & $0)$ & $(0,\pi/2)$ \\
29 & $(1800,2850)$ & $(1430,2200)$ & $(0$ & $1)$ & $(-1$ & $0)$ & $(\pi/2,\pi)$ \\
30 & $(1800,2850)$ & $(1430,2200)$ & $(0$ & $1)$ & $(0$ & $1)$ & $(0,\pi/2)$ \\
31 & $(1800,2850)$ & $(1430,2200)$ & $(0$ & $1)$ & $(0$ & $1)$ & $(\pi/2,\pi)$ \\
32 & $(1800,2850)$ & $(2200,3600)$ & $(-1$ & $0)$ & $(-1$ & $0)$ & $(0,\pi/2)$ \\
33 & $(1800,2850)$ & $(2200,3600)$ & $(-1$ & $0)$ & $(-1$ & $0)$ & $(\pi/2,\pi)$ \\
34 & $(1800,2850)$ & $(2200,3600)$ & $(-1$ & $0)$ & $(0$ & $1)$ & $(0,\pi/2)$ \\
35 & $(1800,2850)$ & $(2200,3600)$ & $(-1$ & $0)$ & $(0$ & $1)$ & $(\pi/2,\pi)$ \\
36 & $(1800,2850)$ & $(2200,3600)$ & $(0$ & $1)$ & $(-1$ & $0)$ & $(0,\pi/2)$ \\
37 & $(1800,2850)$ & $(2200,3600)$ & $(0$ & $1)$ & $(-1$ & $0)$ & $(\pi/2,\pi)$ \\
38 & $(1800,2850)$ & $(2200,3600)$ & $(0$ & $1)$ & $(0$ & $1)$ & $(0,\pi/2)$ \\
39 & $(1800,2850)$ & $(2200,3600)$ & $(0$ & $1)$ & $(0$ & $1)$ & $(\pi/2,\pi)$ \\
\hline
\end{tabular}

\label{tab:schemeconf2}
\end{table}

\begin{table}
\caption{Measurements of $a^{\hat{T}\text{-odd}}_{CP}$ and $a^{\hat{T}\text{-odd}}_{P}$ in specific phase-space regions for the
 \mbox{$B^{0}\to p\bar{p} K^{+}\pi^{-}$} decay for binning scheme B. Each value is obtained through an independent fit to the $B^{0}$ invariant-mass distribution of the candidates in the corresponding region of the phase space.}
\centering
\begin{tabular}{c r@{\,$\pm$\,}l r@{\,$\pm$\,}l r@{\,$\pm$\,}l r@{\,$\pm$\,}l}
Region & \multicolumn{2}{c}{$A_{\hat{T}}(\%)$} & \multicolumn{2}{c}{$\bar{A}_{\hat{T}}(\%)$} & \multicolumn{2}{c}{$a^{\hat{T}\text{-odd}}_{CP}(\%)$} & \multicolumn{2}{c}{$a^{\hat{T}\text{-odd}}_{P}(\%)$} \\\hline
0 & $-26.7$&17.8 & $-21.9$&12.9 & $-2.4$&11.0 & $-24.3$&11.0\\1 & 5.4&15.8 & $-1.6$&20.7 & 3.5&13.0 & 1.9&13.0 \\2 & $-7.3$&11.1 & 18.9&17.4 & $-13.1$&10.3 & 5.8&10.3 \\3 & 15.4&12.8 & 5.0&13.7 & 5.2&9.4 & 10.2&9.4 \\4 & $-21.9$&13.9 & 26.1&16.3 & $-24.0$&10.7 & 2.1&10.7 \\5 & $-13.4$&13.9 & 21.9&22.3 & $-17.6$&13.1 & 4.2&13.1 \\6 & $-19.3$&10.4 & $-15.3$&11.4 & $-2.0$&7.7 & $-17.3$&7.7 \\7 & 0.7&10.9 & 2.8&8.4 & $-1.1$&6.9 & 1.8&6.9 \\8 & $-5.1$&12.8 & $-8.0$&13.2 & 1.5&9.2 & $-6.5$&9.2 \\9 & 6.6&5.8 & 4.0&5.6 & 1.3&4.0 & 5.3&4.0 \\10 & 0.7&9.0 & 30.2&12.2 & $-14.8$&7.6 & 15.4&7.6 \\11 & 30.9&8.7 & 0.2&9.4 & 15.3&6.4 & 15.6&6.4 \\12 & 38.4&16.8 & 22.7&10.7 & 7.9&10.0 & 30.58&9.96\\13 & 11.6&10.2 & 14.2&8.8 & $-1.3$&6.7 & 12.9&6.7 \\14 & $-24.1$&10.5 & 6.4&13.2 & $-15.3$&8.4 & $-8.8$&8.4 \\15 & $-18.8$&8.6 & 10.2&9.5 & $-14.5$&6.4 & $-4.3$&6.4 \\16 & 7.3&11.9 & $-20.9$&15.6 & 14.1&9.8 & $-6.8$&9.8 \\17 & $-2.6$&10.4 & $-1.5$&9.0 & $-0.5$&6.9 & $-2.1$&6.9 \\18 & 10.5&11.7 & 24.0&12.6 & $-6.8$&8.6 & 17.2&8.6 \\19 & 1.1&10.9 & 20.5&9.8 & $-9.7$&7.3 & 10.8&7.3 \\20 & 28.6&32.0 & $-28.9$&21.9 & 28.7&19.4 & $-0.1$&19.4 \\21 & 0.4&19.3 & 9.1&15.6 & $-4.4$&12.4 & 4.7&12.4 \\22 & 36.2&19.0 & $-13.9$&19.2 & 25.1&13.4 & 11.1&13.6 \\23 & $-4.3$&12.5 & $-28.1$&14.3 & 11.9&9.5 & $-16.2$&9.5 \\24 & 11.4&5.4 & 9.4&5.1 & 1.0&3.7 & 10.4&3.7 \\25 & 6.1&4.2 & 2.2&4.3 & 1.9&3.0 & 4.2&3.0\\26 & $-9.7$&7.6 & $-20.8$&7.6 & 5.6&5.4 & $-15.3$&5.4\\27 & $-13.1$&5.9 & $-17.7$&5.8 & 2.3&4.1 & $-15.4$&4.1 \\28 & 7.0&9.5 & $-2.2$&10.5 & 4.6&7.1 & 2.4&7.1 \\29 & 14.6&8.9 & 13.3&9.6 & 0.6&6.5 & 13.9&6.5 \\30 & $-6.6$&11.4 & $-21.2$&9.7 & 7.3&7.5 & $-13.9$&7.5\\31 & $-5.3$&6.7 & $-13.0$&8.0 & 3.8&5.2 & $-9.2$&5.2\\32 & 5.5&4.8 & 15.5&4.8 & $-5.0$&3.4 & 10.5&3.4 \\33 & 6.1&3.4 & 5.3&3.4 & 0.4&2.4 & 5.7&2.4 \\34 & $-11.9$&6.7 & $-9.9$&6.5 & $-1.0$&4.7 & $-10.9$&4.7 \\35 & $-0.9$&5.1 & $-3.8$&5.0 & 1.4&3.6 & $-2.3$&3.6\\36 & 5.9&6.3 & $-2.6$&7.1 & 4.2&4.8 & 1.6&4.8 \\37 & 12.4&5.7 & 10.8&5.9 & 0.8&4.1 & 11.6&4.1 \\38 & $-4.6$&7.1 & $-0.5$&6.8 & $-2.0$&4.9 & $-2.5$&4.9 \\39 & 3.3&5.0 & 0.5&5.3 & 1.4&3.6 & 1.9&3.6 \\\hline
\end{tabular}
\label{tab:measured_asymm_combinedB}
\end{table}
\newpage

\section{Background-subtracted phase space distributions}
\label{sec:bkgsubtr}
\begin{figure}[H]
    \centering
    \includegraphics[scale=0.7]{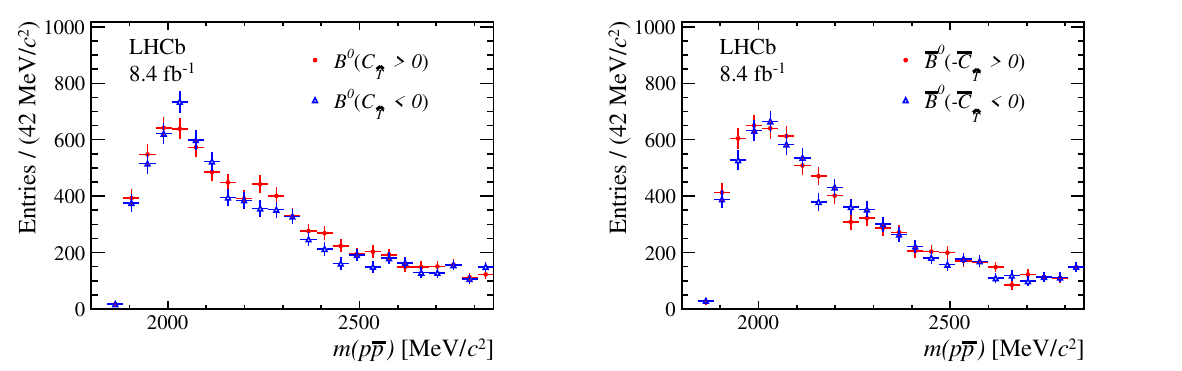}
    \includegraphics[scale=0.7]{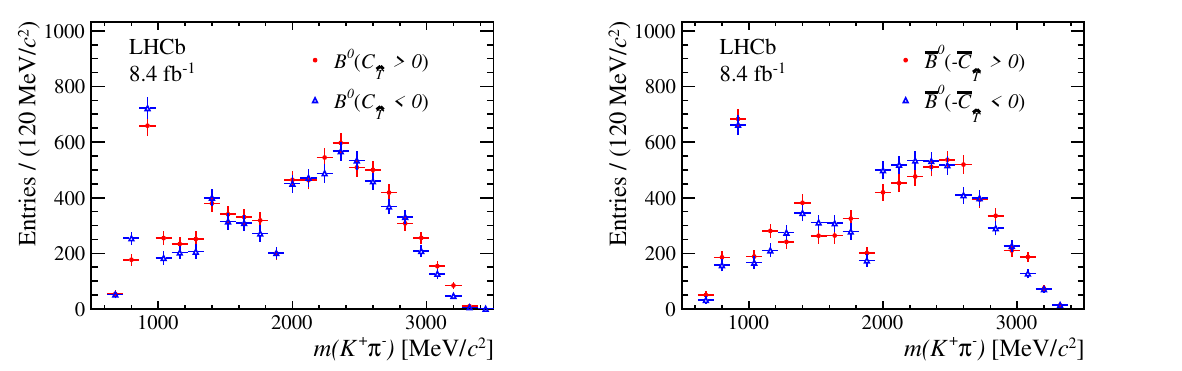}
    \includegraphics[scale=0.7]{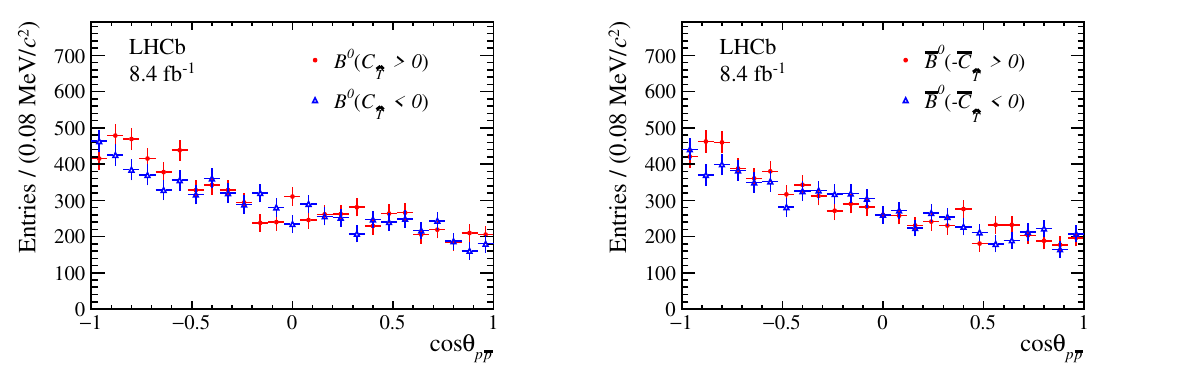}
    \includegraphics[scale=0.7]{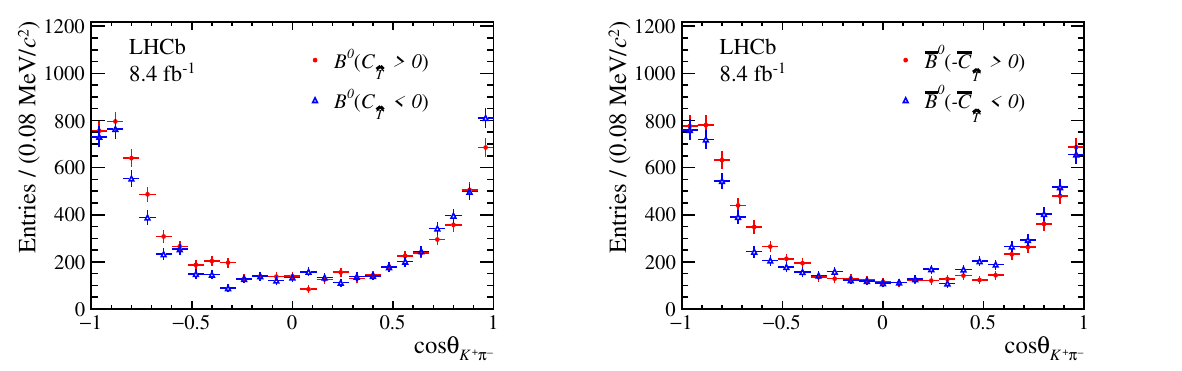}
    \includegraphics[scale=0.7]{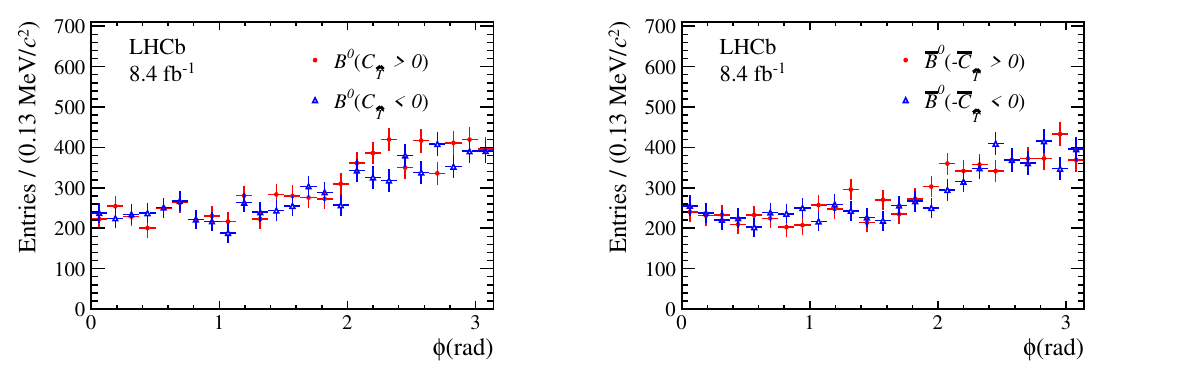}
    \caption{Background-subtracted distributions of $B^{0}$ ($\bar{B}^{0}$) candidates in the variables $m_{p
  \bar p}$, $m_{K^{+}\pi^{-}}$, $\cos\theta_{p
  \bar p}$, $\cos\theta_{K^{+}\pi^{-}}$ and $\phi$ with $C_{\hat{T}}>0$ and $C_{\hat{T}}<0$ ($-\bar{C}_{\hat{T}}>0$ and $-\bar{C}_{\hat{T}}<0$).}
    \label{fig:background_subracted_phs_Run1}
\end{figure}

The background-subtracted distributions of $m_{p
  \bar p}$, $m_{K^{+}\pi^{-}}$, $\cos\theta_{p
  \bar p}$, $\cos\theta_{K^{+}\pi^{-}}$ and $\phi$ obtained using the $sPlot$ technique~\cite{Pivk:2004ty} for $B^{0}$ ($\bar{B}^{0}$) with $C_{\hat{T}}>0$ and $C_{\hat{T}}<0$  ($-\bar{C}_{\hat{T}}>0$ and $-\bar{C}_{\hat{T}}<0$) of the $B^{0}\to p\bar{p} K^{+}\pi^{-}$ decay are shown in Fig.~\ref{fig:background_subracted_phs_Run1} for the combined data. 

\bibliographystyle{LHCb}
\bibliography{main,standard,LHCb-PAPER,LHCb-CONF,LHCb-DP,LHCb-TDR}

\newpage
\centerline
{\large\bf LHCb collaboration}
\begin
{flushleft}
\small
R.~Aaij$^{32}$\lhcborcid{0000-0003-0533-1952},
A.S.W.~Abdelmotteleb$^{50}$\lhcborcid{0000-0001-7905-0542},
C.~Abellan~Beteta$^{44}$,
F.~Abudin{\'e}n$^{50}$\lhcborcid{0000-0002-6737-3528},
T.~Ackernley$^{54}$\lhcborcid{0000-0002-5951-3498},
B.~Adeva$^{40}$\lhcborcid{0000-0001-9756-3712},
M.~Adinolfi$^{48}$\lhcborcid{0000-0002-1326-1264},
H.~Afsharnia$^{9}$,
C.~Agapopoulou$^{13}$\lhcborcid{0000-0002-2368-0147},
C.A.~Aidala$^{76}$\lhcborcid{0000-0001-9540-4988}
% ...

{\footnotesize \it

% ...
$^{9}$Universit{\'e} Clermont Auvergne, CNRS/IN2P3, LPC, Clermont-Ferrand, France\\
% ...
$^{13}$LPNHE, Sorbonne Universit{\'e}, Paris Diderot Sorbonne Paris Cit{\'e}, CNRS/IN2P3, Paris, France\\
% ...
$^{32}$Nikhef National Institute for Subatomic Physics, Amsterdam, Netherlands\\
% ...
$^{40}$Instituto Galego de F{\'\i}sica de Altas Enerx{\'\i}as (IGFAE), Universidade de Santiago de Compostela, Santiago de Compostela, Spain\\
% ...
$^{44}$Physik-Institut, Universit{\"a}t Z{\"u}rich, Z{\"u}rich, Switzerland\\
% ...
$^{48}$H.H. Wills Physics Laboratory, University of Bristol, Bristol, United Kingdom\\
% ...
$^{50}$Department of Physics, University of Warwick, Coventry, United Kingdom\\
% ...
$^{54}$Oliver Lodge Laboratory, University of Liverpool, Liverpool, United Kingdom\\
% ...
$^{62}$Syracuse University, Syracuse, NY, United States\\
% ...
$^{76}$University of Michigan, Ann Arbor, MI, United States, associated to $^{62}$\\
\bigskip
% ...
}
\end{flushleft} 

\end{document}